\documentclass[fleqn,usenatbib]{mnras}

\usepackage{newtxtext,newtxmath}

\usepackage[T1]{fontenc}

\DeclareRobustCommand{\VAN}[3]{#2}
\let\VANthebibliography\thebibliography
\def\thebibliography{\DeclareRobustCommand{\VAN}[3]{##3}\VANthebibliography}

\usepackage{graphicx}	
\usepackage{amsmath}	
\usepackage{longtable}

\title[\textit{TESS} Exoplanet Occurrence Rates of FGK Dwarfs]{Demographics of Close-In \textit{TESS} Exoplanets Orbiting FGK Main-sequence Stars}

\author[K. Cui et al.]{
Kaiming Cui,$^{1,2}$\thanks{E-mail: Kaiming.Cui@warwick.ac.uk}
David J. Armstrong,$^{1,2}$
Andreas Hadjigeorghiou,$^{1,2}$
Marina Lafarga,$^{1,2}$
\newauthor
Vedad Kunovac,$^{1,2}$
Lauren Doyle,$^{1,2}$
Luis Agust\'in Nieto$^{3,4}$
and Rodrigo F. D\'iaz$^{3,5}$
\\
$^{1}$Department of Physics, University of Warwick, Gibbet Hill Road, Coventry CV4 7AL, UK\\
$^{2}$Centre for Exoplanets and Habitability, University of Warwick, Gibbet Hill Road, Coventry CV4 7AL, UK\\
$^{3}$Instituto de Ciencias F\'isicas (CONICET / ECyT-UNSAM), Campus Miguelete, 25 de Mayo y Francia, (1650) Buenos Aires, Argentina\\
$^{4}$Gerencia de Tecnología de la información y de las Comunicaciones (GTIC), Subgerencia Vinculación y Desarrollo de Nuevas Tecnologías de la Información,\\ DTE-CNEA. Centro Atómico Constituyentes, Av. Gral. Paz 1499, (1650) Buenos Aires, Argentina\\
$^{5}$Instituto Tecnol\'ogico de Buenos Aires (ITBA), Iguaz\'u 341, Buenos Aires, CABA C1437, Argentina
}

\date{Accepted XXX. Received YYY; in original form ZZZ}

\pubyear{\the\year{}}

\begin{document}
\label{firstpage}
\pagerange{\pageref{firstpage}--\pageref{lastpage}}
\maketitle

\begin{abstract}
Understanding the demographics of close-in planets is crucial for insights into exoplanet formation and evolution. We present a detailed analysis of occurrence rates for close-in (0.5--16 day) planets with radii between 2 and {20}\,$R_{\earth}$ around FGK main-sequence stars. Our study uses a comprehensive sample from four years of \textit{TESS} Science Processing Operations Center full-frame image data cross-matched with \textit{Gaia}, analysed through our rigorous detection, vetting, and validation pipeline. Using high-confidence planet candidates, we apply a hierarchical Bayesian model to determine occurrence rates in the two-dimensional orbital period-radius plane.
Our results are presented using 10-by-10 bins across the period-radius parameter space, offering unprecedented resolution and statistical precision. 
We find an overall occurrence rate of $9.4^{+0.7}_{-0.6}\%$. When using identical binning, our occurrence rate posteriors distributions align with \textit{Kepler}'s but have a magnitude smaller uncertainties on average.
{For hot Jupiters, we estimate the overall occurrence rate of $0.39^{+0.03}_{-0.02}\%$. This value is consistent with the previous \textit{Kepler} FGK-type result within $1\sigma$. We find an overall occurrence rate of Neptunian desert planets of $0.08\pm0.01\%$, to our knowledge the first such determination.} Additionally, in a volume-limited \textit{Gaia} subsample within 100\,pc in the same parameter region, we measure an overall planet occurrence rate of $15.4^{+1.6}_{-1.5}\%$ and a hot Jupiter occurrence rate of $0.42^{+0.16}_{-0.12}\%$. 
Our results establishes an improved foundation for constraining theoretical models of exoplanet populations.
\end{abstract}




\section{Introduction}
Over the past three decades, the number of confirmed exoplanets has grown from a handful to several thousand, enabling the transition from individual detections to statistical studies of planetary populations. The statistical characterisation of these populations, often referred to as exoplanet demographics, provides crucial constraints on models of planet formation and evolution \citep[e.g.][]{mordasiniExtrasolarPlanetPopulation2009, muldersExoplanetPopulationObservation2019,drazkowskaPlanetFormationTheory2023}. Large, uniform transit and radial velocity (RV) surveys have revealed striking structures in the period-radius and period-mass planes, such as the “hot Jupiter (HJ)” population \citep[e.g.][]{cummingKeckPlanetSearch2008, wrightFrequencyHotJupiters2012}, the “Neptunian desert” \citep[e.g.][]{szaboShortperiodCensorSubJupiter2011, beaugeEMERGINGTRENDSPERIOD2012, mazehDearthShortperiodNeptunian2016}, and the “radius valley” \citep[e.g.][]{fultonCaliforniaKeplerSurveyIII2017, vaneylenAsteroseismicViewRadius2018}. Each of these features encodes the imprint of processes such as disk migration, atmospheric escape, and core accretion efficiency \citep[e.g.][]{dawsonOriginsHotJupiters2018,owenPhotoevaporationHigheccentricityMigration2018,owenEvaporationValleyKepler2017, lopezROLECOREMASS2013,jinPLANETARYPOPULATIONSYNTHESIS2014}, and mapping them with increasing precision remains a central goal of exoplanetary science.

The NASA \textit{Kepler} mission \citep{boruckiKeplerPlanetDetectionMission2010} offers the first opportunity for statistically robust measurements of planetary occurrence rates across a wide range of orbital periods and planetary radii \citep[e.g.][]{howardPLANETOCCURRENCE0252012, fressinFALSEPOSITIVERATE2013, dressingOccurrencePotentiallyHabitable2015, petiguraCaliforniaKeplerSurveyIV2018, hsuOccurrenceRatesPlanets2019, hsuOccurrenceRatesPlanets2020,kunimotoSearchingEntiretyKepler2020,brysonReliabilityCorrectionKey2020,dattiloUnifiedTreatmentKepler2023}. Its unprecedented photometric precision reveals not only the abundance of small planets but also the fine structures in their distribution \citep[e.g.,][]{fultonCaliforniaKeplerSurveyIII2017,dattiloUnifiedTreatmentKepler2024}. However, \textit{Kepler}’s focus on a single, narrow field restricted its stellar sample to relatively faint targets, making follow-up and confirmation challenging. 

NASA’s Transiting Exoplanet Survey Satellite \citep[\textit{TESS};][]{rickerTransitingExoplanetSurvey2014} enables the extension of the demographic revolution to bright, nearby stars across the entire sky. \textit{TESS} observes the sky in a series of overlapping 27 day sectors, with extended coverage in its continuous viewing zones, enabling sensitive searches for short-period ($P \lesssim 10$--20\,days) transiting planets \citep[e.g.][]{guerreroTESSObjectsInterest2021}. Its all-sky coverage, emphasis on bright stars, and systematic observing strategy make it particularly powerful for constraining the occurrence rates of close-in planets. For FGK main-sequence stars, the brightness of \textit{TESS} targets allows precise stellar characterisation \citep[e.g.][]{stassunTESSInputCatalog2018, stassunRevisedTESSInput2019} and facilitates RV follow-up \citep[e.g.][]{huangTESSDiscoveryTransiting2018}, yielding accurate planetary masses and densities.

The \textit{TESS} mission has already yielded several demographic studies of short-period planets \citep[e.g.,][]{fernandesPterodactylsToolUniformly2022,mentOccurrenceRateTerrestrial2023,bryantOccurrenceRateGiant2023,vachOccurrenceSmallShortperiod2024,meltonDIAmanteTESSAutoRegressive2024b}. Most such studies have focused on young stars or M dwarfs, or have analysed early subsets of the \textit{TESS} dataset. Some have revealed hints of differences between the \textit{TESS} and \textit{Kepler} planet populations, likely due to differences in target selection, magnitude limits, and observing baselines. A comprehensive, statistically robust census of close-in planets around full-sky FGK main-sequence stars in the \textit{TESS} era is still lacking. Such a study can test the universality of features such as refine the boundaries of the Neptunian desert, improve the statistical precision of hot Jupiter occurrence rates, and provide a uniform, bright-star sample optimised for follow-up.

In this work, we present a systematic demographic analysis of close-in ($0.5 < P < 16$ days) \textit{TESS} planets orbiting FGK main-sequence stars. We describe our sample selection in Section~\ref{sec:sample} and our completeness estimation in Section~\ref{sec:methods}, and outline our occurrence-rate methodology in Section~\ref{subsec:occ-method}. We then present occurrence rates for our stellar sample (Section~\ref{subsec:TESS occurrence rate}), a Gaia 100\,pc volume-limited sample (Section~\ref{subsec:100pc}), for hot Jupiters, {and Neptunian desert} (Section~\ref{subsec:HJ-occ}). Next, we discuss the impact of different models, thresholds {and RUWE values} on our occurrence rate results (Section~\ref{subsec:thresholds effect} {and Section~\ref{subsec:RUWE-effect}}) and compare our findings with those from Kepler (Section~\ref{subsec:TESS-Kepler-occ-comp}). Finally, we offer future prospects in Section~\ref{subsec:future}.

\section{Stellar sample}\label{sec:sample}
This study aims to determine the planet occurrence rate from the full-sky \textit{TESS} observations of FGK-type main sequence stars. To achieve this, we use a main sequence sample from \citet{doyleTESSSPOCFFITarget2024}, which cross-matches objects from the \textit{TESS}-Science Processing Operations Center (SPOC) full-frame image (FFI) releases \citep{caldwellTESSScienceProcessing2020} with data from \textit{Gaia} Data Release 2 \citep[][]{brownGaiaDataRelease2018} and Data Release 3 \citep[][]{smartGaiaEarlyData2021}. The \textit{TESS}-SPOC observations are from the first four years (Sectors 1--55), which cover most of the fields in the Northern and Southern hemispheres and the ecliptic region. The \textit{TESS}-SPOC select sources mainly by FFI field stars with Tmag < 13.5, but also include some fainter high value objects. Thus we require that the stars in our sample must be brighter than Gmag 14, have a {\tt parallax\_over\_error} > 5, and surface gravity $\log g > 3.5$ to ensure that they are main sequence stars. Full details of the sample construction and its properties are available in \citet{doyleTESSSPOCFFITarget2024} and Lafarga et al. (submitted).

\begin{figure}
    \centering
    \includegraphics[width=\columnwidth]{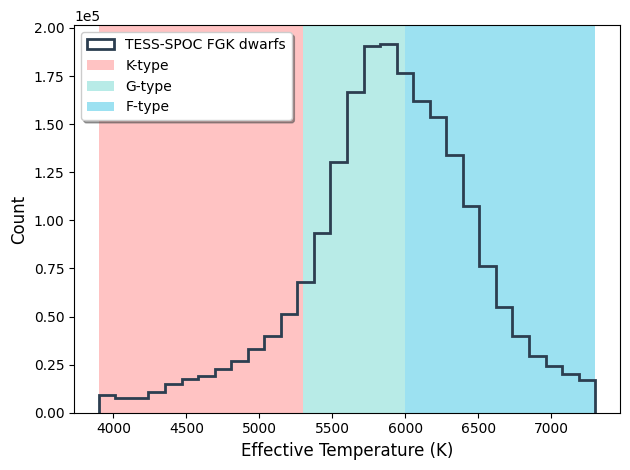}
    \caption{Effective temperature distributions of our TESS-SPOC FGK main sequence stars. Shaded regions represent the temperature boundaries for different FGK spectral types.}
    \label{fig:stellar-temperature}
\end{figure}

From the above sample, we further select FGK-type stars using the \textit{TESS} Input Catalogue v8.2 \citep[TIC;][]{stassunRevisedTESSInput2019,paegertTESSInputCatalog2021}, defined by an effective-temperature range of 3900--7300\,K \citep{pecautINTRINSICCOLORSTEMPERATURES2013,kunimotoComparingApproximateBayesian2020}, yielding a stellar sample of 2,097,359 targets for our detection pipeline. Although not severe, the \textit{TESS}-SPOC target selection is heterogeneous to some extent. Given the sample’s large size and our additional cuts, the resulting targets can be approximated as a uniform, magnitude-limited sample of FGK main-sequence stars. Our planet detection and occurrence rate estimation are established on this sample, and its effective temperature distribution is shown in Fig.~\ref{fig:stellar-temperature}. We download the FFI light curves of our sample from the Mikulski Archive for Space Telescopes \footnote{\url{https://archive.stsci.edu/hlsp/tess-spoc}}. The pre-search data conditioning simple aperture photometry (PDCSAP) is used for our planet detection.

\section{Methods}\label{sec:methods}
The detected planets are characterized by their intrinsic occurrence rate and discovery completeness. To accurately estimate the occurrence rate, it is necessary to assess the efficiency, or completeness, of the selection process. Completeness maps should faithfully represent the entire sequence from observation to validation. We achieve this via simulation, applying the same pipeline to both simulated and observed stellar samples. We will briefly introduce the key procedures and our approach to modelling completeness estimates.

Once we understand the completeness of each stage, calculating the occurrence rate typically involves working backwards. In this study, we implement two algorithms: one for qualitative demonstration and another for quantitative analysis. They can validate each other, offering a comprehensive view with different resolutions and robustnesses.

\subsection{Simulated Sample}\label{subsec:simulated}
The entire process and details of our simulation are described in \citet{hadjigeorghiouRAVENRAnkingValidation2025}. Here, we will introduce the key ideas and choices.

Our simulated sample is built by constructing planetary or binary systems on a target stellar sample. 
The target stellar sample should consist of single stars and have stellar properties that are identically distributed with respect to those of the observed sample. To do so, we also begin with the stellar sample from \citet[][as described in Section~\ref{sec:sample}]{doyleTESSSPOCFFITarget2024}. Differently from the stellar sample for detection and occurrence rate estimation, we exclude \textit{TESS} Objects of Interest (TOIs), community candidates (CTOIs), \textit{Kepler} Objects of Interest (KOIs), and K2 candidates listed in ExoFOP\footnote{\url{https://exofop.ipac.caltech.edu/tess/}}, as well as stars with Gaia Renormalised Unit Weight Error (RUWE) greater than 1.05, ensuring that the remaining targets are likely single stars without known planetary signals. {Note that as part of our false positive simulations, higher RUWE stars are in effect `created' from this sample, so our results are not limited to low RUWE stars.}

Given the scope of our study, we restrict their effective temperature range, based on TIC data, from 3,000 K to 10,000 K. This range is broader than that of our current sample used for occurrence rate calculations, as we are reserving it for future plans. {However, for our occurrence rate calculations and completeness map estimations in this work, we limit both our simulated and observed stellar samples to the same Teff range as our stellar sample (i.e., FGK stars). This ensures that the simulated sample is directly comparable to the stellar sample used in our analysis.} In the end, about 1.2 million \textit{TESS}-SPOC FFI stars are used as the target sample for our simulation. This number is significantly lower than the stellar sample in Section~\ref{sec:sample}, primarily due to the stringent RUWE criterion we applied excluding almost half of the stars. {We will discuss the impact of RUWE on our occurrence rate results in Section \ref{subsec:RUWE-effect}.}

The construction of planetary systems around the target stars is based on orbital simulations and a light curve generator. The planets' orbital parameters are sampled from pre-determined distributions. For the orbital period and planetary radius, we have two different distributions. The first distribution is based on planetary occurrence rates, sampling orbital periods and planet radii according to their relative occurrence rates from \textit{Kepler} results \citep{hsuOccurrenceRatesPlanets2019}. The second distribution is a log-uniform distribution for both period and radius. Both distributions cover orbital periods from 0.5 to 16 days and planet radii from 1 to 16\,$R_\text{\earth}$. We call the sample with parameters drawn based on literature occurrence rates the non-uniform sample, and the sample with uniformly drawn parameters the uniform sample. 

For the other parameters, inclinations are drawn from an isotropic distribution, $p(i)\propto \sin i$ for $i\in[0,\uppi/2]$. The argument of periastron is drawn from a uniform distribution, $\mathcal{U}(0,2\uppi)$. Eccentricities follow the distribution from \citet{moeMindYourPs2017}, with $p_e\propto e^\eta$ and $\eta = -0.3$, truncated at $e_\mathrm{max}(P)$, a function of orbital period defined in Equation (3) of \citet{moeMindYourPs2017}. They are set to zero for periods shorter than 2 days. Albedos are drawn uniformly over $[0, 1]$. Planet masses are assigned using the empirical mass-radius relation of \citet{mullerMassradiusRelationExoplanets2024}. These distributions are applied identically to both the uniform and non-uniform samples. {We also simulate another independent uniform planetary sample with an increased radius range up to 20\,$R_\text{\earth}$; this is reserved only for completeness estimation.}

We also simulate samples for eight different astrophysical false positive (FP) scenarios: 
\begin{enumerate}
    \item hierarchical transiting planet (HTP; a planet transits an unresolved stellar companion of the target star)
    \item nearby transiting planet (NTP; a diluting transit system not associated with the target star)
    \item background transiting planet (BTP; a background transit system blend with the target star)
    \item eclipsing binary (EB; a secondary star eclipses the target star)
    \item background eclipsing binary (BEB; a background EB system blend with the target star)
    \item nearby eclipsing binary (NEB; a diluting EB system not associated with the target star)
    \item hierarchical EB system (HEB; a wide orbit EB system bound to the target star)
    \item nearby hierarchical EB system (NHEB; a diluting HEB system not associated with the target star). 
\end{enumerate}
Full details of the FP parameter configuration are provided in \citet{hadjigeorghiouRAVENRAnkingValidation2025}. We generate light curves for all simulated systems using \texttt{JKTEBOP} \citep[][]{popperPhotometricOrbitsSeven1981,etzelSimpleSynthesisMethod1981,nelsonEclipsingBinarySolutionsSequential1972,southworthEclipsingBinariesObserved2007}. Consequently, not all simulated systems produce a light curve; this is primarily due to non-transiting orbital configurations, though unrealistic stellar and planetary properties can also cause failures. We also exclude systems whose generated light-curve depth is below 300\,ppm to improve efficiency. We omit the BTP scenario in our study because our simulations show an extremely low success rate (fewer than 500 per million), indicating that such cases are rare and can be safely ignored in our validation. Similar to the planetary simulation, the planetary systems in our FPs (HTP, NTP, BTP) are also simulated with orbital parameters drawn from uniform distributions and from literature. 

The simulations are performed using our modified {\tt PASTIS} code\footnote{\url{https://github.com/ckm3/pastis-dev}} \citep[][]{diazPastisBayesianExtrasolar2014,santernePastisBayesianExtrasolar2015}. We simulate at least 30,000 objects with successfully generated light curves for each scenario, and most scenarios involve more than 100,000 objects. The final simulated light curves are created by injecting the generated transiting or eclipsing light curves into the corresponding target stars’ TESS-SPOC light curves. 

Unlike some previous work \citep[e.g.,][]{giacaloneVetting384TESS2020}, we do not include period aliases (e.g., twice or half the orbital period) directly in our simulations; however, they appear in our training samples in certain scenarios. We discuss them in Section~\ref{subsec:vetting&validation}.
Additionally, we have a non-simulated false positive (NSFP) scenario. It represents any stellar variability or noise features not well modelled in other simulated scenarios. Unlike the other scenarios, the NSFP sample is generated from a randomly selected sample of box least square (BLS) peaks on randomly chosen TESS-SPOC light curves after removing TOIs.

\subsection{Observational bias}
Not all planetary systems are observable via the transit method. We refer to the observability of transiting planets in our target sample as an observational bias or observational completeness.

With our {reserved} uniform planet sample, we can directly examine the observational bias of the target stars in orbital period and planet radius space. We first limit the effective temperature of the target stars to FGK spectral types. For each bin in period and radius space, we estimate the yield rate by counting the number of successfully generated transiting planets over the total number simulated. Due to the photometric precision of \textit{TESS} and the reliability requirements for our occurrence rates, we restrict our observational-bias calculations to radii $\geq 2\,R_\text{\earth}$. For consistency, our other completeness calculations use the same parameter-space ranges.

\begin{figure*}
    \centering
    \includegraphics[width=\textwidth]{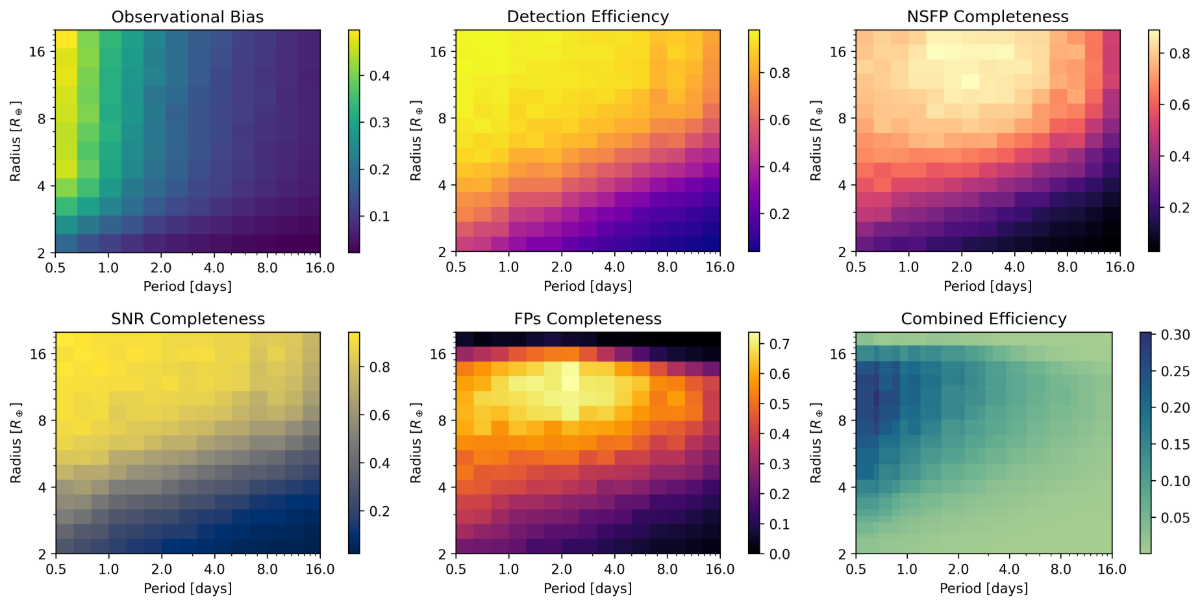}
    \caption{Observational bias, detection efficiency, NSFP completeness, signal-to-noise ratio (SNR) completeness, FP completeness, and combined efficiency of transiting planets in our stellar sample as functions of orbital period and planet radius. Each subplot title describes its process; higher values indicate more planets retained.}
    \label{fig:completeness-grid}
\end{figure*}

The upper-left subplot of Fig.~\ref{fig:completeness-grid} shows the observational bias map for our target stars. Higher completeness values are concentrated at shorter orbital periods and larger planetary radii, primarily reflecting the geometric transit probability in our stellar sample. We also require a minimum candidate transit depth of 300\,ppm to ensure that candidate signals are sufficiently deep to be robustly detectable in our data.

\subsection{Detection efficiency}\label{subsec:detection}
Detection efficiency quantifies the completeness of transiting planets that are successfully detected. We first describe our detection pipeline.

We begin by detrending the light curves using the Savitzky-Golay filter \citep{savitzkySmoothingDifferentiationData1964}, applying a third-degree moving polynomial with a four-day window length. Then, we apply a GPU-based Box-Least Squares (BLS) algorithm using {\tt cuvarbase}\footnote{\url{https://github.com/johnh2o2/cuvarbase}} to the entire light curve to identify the five most significant peaks for each stellar target. We further select candidates with a signal detection efficiency \citep[SDE;][]{kovacsBoxfittingAlgorithmSearch2002} greater than 7 and a multiple event statistic (MES) greater than 0.8. Our MES algorithm is a modified implementation of \citet{jenkinsTESSScienceProcessing2016}; see \citet{hadjigeorghiouRAVENRAnkingValidation2025} for details. Selected candidates are further excluded if their depth varies by more than 50\% from the average across sectors or if the depth in any sector is below 200 ppm. Note this is a separate sector-by-sector cut, as opposed to our 300\,ppm average depth limit in our simulation.

Analogous to our treatment of observational bias, we compute the detection efficiency in the orbital period-planet radius plane. In each bin, we calculate the fraction of correctly recovered planets (period difference within 2\%, epoch difference within 0.5 days, and {detected-to-injected transit-depth ratio and its reciprocal must be less than 3}) relative to the number injected. The top-middle subplot of Fig.~\ref{fig:completeness-grid} shows the detection efficiency estimated from our simulated sample. Generally, shorter periods and larger radii yield higher detection efficiency, consistent with previous signal-to-noise ratio (SNR)-based estimates \citep[e.g.,][]{muldersExoplanetPopulationObservation2018}.

\subsection{Vetting and Validation Completeness}\label{subsec:vetting&validation}
Although many planetary candidates pass the thresholds after the BLS process, a significant number of false positives remain. To reliably identify true exoplanets and rule out signals caused by instrumental effects or astrophysical phenomena, further vetting and validation are usually required \citep[e.g.,][]{mortonFALSEPOSITIVEPROBABILITIES2016,thompsonPlanetaryCandidatesObserved2018,giacaloneVetting384TESS2020,armstrongExoplanetValidationMachine2021,valizadeganExoMinerHighlyAccurate2022}. In this work, vetting and validation are carried out by our \texttt{RAVEN} pipeline\footnote{\url{https://github.com/ahadjigeorghiou/RAVEN-Pipeline}} \citep[][]{hadjigeorghiouPositionalProbabilityTrue2024,hadjigeorghiouRAVENRAnkingValidation2025}, which is designed for \textit{TESS} exoplanet candidates. Here, we provide only a brief overview of the pipeline; for more details, please refer to \citet{hadjigeorghiouRAVENRAnkingValidation2025}.

\texttt{RAVEN} uses two different feature-based machine learning classifiers for its vetting and validation, specifically a Gradient Boosted Decision Tree (GBDT) based on the XGBoost \citep{chenXGBoostScalableTree2016} and a Gaussian process (GP) classifier \citep{rasmussenGaussianProcessesMachine2008}. The classifiers are binary and trained to distinguish true planetary signals against each astrophysical FP scenario and the NSFP. Due to the uniqueness of the latter, the classification result is only used for vetting the candidates. All classifiers are trained on our simulated dataset, which is split into training, validation, and test sets in an 80\%, 10\%, and 10\% ratio, respectively. The features used by the classifiers generally fall into five categories: stellar properties (e.g., BP-RP, distance), eclipse parameters (e.g., transit depth, duration, and ratio of semi-major axis and planetary radius), eclipse significance (e.g., SNR, MES, SES), nearby metrics (target fraction and nearby fraction, used only for nearby scenarios), and Self-Organising Map \citep[][]{kohonenSelforganizedFormationTopologically1982,armstrongTransitShapesSelforganizing2017} features. A complete list of features is available in \citet{hadjigeorghiouRAVENRAnkingValidation2025}.

The feature data may contain orbital period aliases. We exclude these aliases from our planetary scenarios because aliased planets are not considered true detections. However, we include the aliases in our EB-based scenarios, as they represent a significant subset of common FPs. Therefore, since our model lacks a training set for aliased planets, our final planet candidates require additional alias removal (see Section~\ref{subsec:planet candidates}).

Given the presence of seven different FP scenarios, we train separate classifiers for each scenario to distinguish FPs from planets. Including NSFPs, we thus have eight binary classifiers, with each classifier utilizing two machine learning models, resulting in a total of 16 models. To ensure reliable statistical results, we also balance the sample sizes so that, for each FP and planet pair, the training set contains an equal number of FPs and planets. We use the average score of the GBDT and GP models as the combined score. The classification precision on our test set exceeds 0.99 across all scenarios, while recalls are above 0.85 in all cases except for NTP (0.78) and HTP (0.80).

We also evaluate the calibration performance, which generally follows the perfect calibration line \citep[see Fig. 13 in][]{hadjigeorghiouRAVENRAnkingValidation2025}. Therefore, we interpret the model-predicted scores as probabilities for each scenario. Then, we compute posterior probabilities by combining these scenario probabilities with priors for each candidate under Bayes’ theorem. The prior for a candidate includes the occurrence rate, detection rate, and positional probability \citep[see][]{hadjigeorghiouPositionalProbabilityTrue2024}. Thus, vetting and validation here can be achieved by setting appropriate posterior probability thresholds to rule out FPs and NSFPs.


Due to the different planetary signal injections (Section~\ref{subsec:simulated}), we have two simulated planetary samples: one uniform and one non-uniform. Accordingly, we train two models per classifier for each planetary FP---one on the uniform sample and one on the non-uniform sample. As a result, we obtain two posterior probabilities for each object, depending on whether the uniform or non-uniform model is used. The differing orbital period and radius distributions influence each model to learn the respective intra-distribution of its sample. Comparing these two models provides a test of the robustness of our occurrence rate estimation.

In this work, we set the thresholds for the NSFP and FP classifiers to 0.9 (where 0 corresponds to 100\% false positives and 1 to 100\% planets) for both non-uniform and uniform models. We selected this value because model calibration is more reliable for scores above 0.9 \citep{hadjigeorghiouRAVENRAnkingValidation2025}. Additionally, excessively high thresholds would result in fewer candidates, causing completeness estimation to fail due to numerous zero detections, as well as increased Poisson noise in our occurrence rate calculation.

\begin{figure*}
    \centering
    \includegraphics[width=\linewidth]{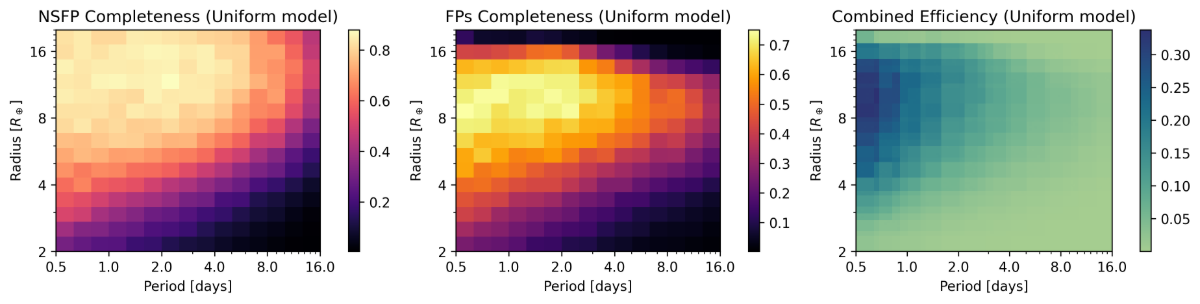}
    \caption{NSFP completeness, FPs completeness, and combined efficiency as a function of the orbital period and planet radius. The NSFP and FPs models are trained on our uniformly injected sample. }
    \label{fig:completeness-grid-uniform}
\end{figure*}

Similar to the detection efficiency, we estimate our NSFP and FPs completeness using the simulated {planet sample}. Specifically, in each bin, we compute the fraction of correctly classified planets relative to the number injected. For FPs, a planet candidate is counted as correctly classified only if all seven models have posterior probabilities greater than 0.9. The top-right and bottom-middle subplots of Fig.~\ref{fig:completeness-grid} illustrate the completeness of the NSFP and FP with non-uniform models. We observe lower completeness in the FP for regions with larger radii and longer periods, where candidates can resemble EBs. In areas with smaller radii and shorter periods, the model becomes confused by scenarios such as BEB, NTP, and other nearby cases. Similarly, Fig.~\ref{fig:completeness-grid-uniform} shows the completeness of the NSFP and FP classifiers trained on the uniform sample. Because the uniformly simulated planet distribution only reflects observational bias, the FP and NSFP classifiers learn this distribution, resulting in higher completeness in regions where planets are more abundant.

Additionally, we introduce a SNR cut to further eliminate weak signals, particularly those resulting from misclassified NSFPs. Our SNR is estimated from the trapezoidal fit results, and we apply a threshold of 10. The bottom-left subplot of Fig.~\ref{fig:completeness-grid} shows the resulting completeness map after our injection-and-recovery tests on the simulated planets. The SNR completeness shows the expected smoothly increasing completeness towards shorter period and larger radius.

Different thresholds result in different model outputs and corresponding completeness maps, which in turn affect the occurrence rates. We discuss the impact of varying thresholds further in Section~\ref{subsec:thresholds effect}.

\subsection{Occurrence rate estimation}\label{subsec:occ-method}
We apply two commonly used methods to estimate the occurrence rate: the inverse detection efficiency method \citep[IDEM; e.g.,][]{howardPLANETOCCURRENCE0252012,dattiloUnifiedTreatmentKepler2023} and a hierarchical Bayesian model of the Poisson process \citep[e.g.,][]{foreman-mackeyEXOPLANETPOPULATIONINFERENCE2014}.

\subsubsection{Inverse detection efficiency method}\label{subsubsec:IDEM}
The occurrence rate of IDEM in a specific bin, denoted as $\Gamma$, can be generalized using the following equation:
\begin{equation}
    \centering
    \Gamma = \frac{N_{\rm p}}{N_*} \frac{\langle r \rangle}{\langle q \rangle},
\end{equation}
where $N_{\rm p}$ represents the number of planets within a specific bin, $N_*$ denotes the number of host stars, $\langle r \rangle$ is the average reliability for that bin, and $\langle q \rangle$ is the average completeness for that bin. In most cases, the $\langle r \rangle$ is one as they are vetted or confirmed as planets by different observations.
However, IDEM is not well-established in probability theory and can lead to biased results because it does not adequately account for uncertainties in transit parameters. Also, when detection efficiency is low or detections are few or absent, it leads to an underestimation and significant fluctuations in the occurrence rate \citep[e.g.,][]{foreman-mackeyEXOPLANETPOPULATIONINFERENCE2014,hsuImprovingAccuracyPlanet2018,zhuExoplanetStatisticsTheoretical2021}.

We demonstrate the use of IDEM with a kernel density function (KDE) to qualitatively measure planet occurrence, known as detection efficiency weighted KDE (wKDE). This method has been applied in various studies \citep[e.g.,][]{mortonRADIUSDISTRIBUTIONPLANETS2014,petiguraCaliforniaKeplerSurveyRadius2022, dattiloUnifiedTreatmentKepler2023} to provide a smooth distribution without binning artifacts. We compute the wKDE by applying a KDE with weights $w = r/q$. Our reliability uses the RAVEN probability (the minimum of FP probabilities and NSFP probability). Our completeness is modelled using Beta distributions with the Jeffreys prior, based on our injection and recovery tests \citep[$\alpha = 1/2 + n_\mathrm{rec}, \beta = 1/2 + n_{\rm inj} - n_{\rm rec}$;][]{jeffreysInvariantFormPrior1946}. The bottom right subplots of Fig.~\ref{fig:completeness-grid} and Fig.~\ref{fig:completeness-grid-uniform} display the combined efficiency for the non-uniform and uniform models. {This is the product of observational bias and the completeness of our pipeline, where completeness is measured as the number of simulated planets that pass through the whole pipeline divided by the total number of injected planets.}

\subsubsection{Hierarchical Bayesian method}\label{subsubsec:bayesian}
To reliably quantify the occurrence rate, we use the hierarchical Bayesian methodology to model the underlying Poisson process \citep[e.g.,][]{youdinExoplanetCensusGeneral2011,tabachnikMaximumlikelihoodMethodEstimating2002,foreman-mackeyEXOPLANETPOPULATIONINFERENCE2014,fultonCaliforniaLegacySurvey2021}. Compared to IDEM, this approach provides robust uncertainty estimations and is more reliable for cases with low detections and low efficiencies. This is particularly important for quantifying the Neptunian desert region. In this work, we slightly modify the original form to include the uncertainties of completenesses, and the reliabilities of our candidates. The log-likelihood can be written as 

\begin{equation}
\begin{aligned}
    \ln \mathcal{L}(\boldsymbol{\theta}) = & - \int Q(\boldsymbol{\omega}) \Gamma(\boldsymbol{\omega} | \boldsymbol{\theta}) \, \mathrm{d}\boldsymbol{\omega} \\
    & + \sum_{k=1}^{K} r_k\cdot\ln\left( \frac{1}{N_k} \sum_{n=1}^{N_k} \frac{Q(\boldsymbol{\omega}_k^{(n)})\Gamma(\boldsymbol{\omega}_k^{(n)} | \boldsymbol{\theta})}{p(\boldsymbol{\omega}_k^{(n)})}\right).
\end{aligned}
\end{equation}

Here, $\boldsymbol{\theta}$ are the parameters for the occurrence rate model. $\boldsymbol{\omega}$ represents physical parameters (orbital period $P$ and planet radius $R_\mathrm{p}$ in this study). Thus $\Gamma(\boldsymbol{\omega})$ is the occurrence rate density, defined as $\Gamma = {\mathrm{d}^2N_\mathrm{p}}/\left({\mathrm{d}\ln{P}\ \mathrm{d}\ \ln{R_\mathrm{p}}}\right)$. $Q(\boldsymbol{\omega})$ is the completeness map, while $r_k$ denotes the reliability of the $k$-th planet. $N_k$ is the number of posterior samples of the $k$-th planet's physical parameters. The term $\boldsymbol{\omega}_k^{(n)}$ refers to the $n$-th sample for the same $k$-th planet. Additionally, $p(\boldsymbol{\omega}_k^{(n)})$ indicates each parameter's prior probability distribution for every companion's posteriors. Here, $r_k$ serves as a Bernoulli weight on the log-likelihood. When $r_k = 0$, the contribution from the $k$-th planet is omitted; when $r_k = 1$, the log-likelihood reduces to its standard form \citep[][]{foreman-mackeyEXOPLANETPOPULATIONINFERENCE2014}.

To account for uncertainties in completeness, we also sample the completeness maps from Beta distributions: $Q(\boldsymbol{\omega}) \sim \text{Beta}(\boldsymbol{\alpha}, \boldsymbol{\beta})$, as described in Section \ref{subsubsec:IDEM}. We assign the \texttt{RAVEN} probability of each candidate as $r_k$, consistent with IDEM. The posteriors of the planets' physical parameters are fitted and sampled using {\tt juliet} \citep{espinozaJulietVersatileModelling2019}, with normal distributions as priors $p(\boldsymbol{\omega}_k^{(n)})$. These normal distributions use the fitted BLS results as the means and set the standard deviations to 0.1 for both $P$ and $R_\mathrm{p}$. Detailed light curve processing and modelling fitting is available in Section~\ref{subsec:planet candidates}.

We then perform Markov Chain Monte Carlo (MCMC) sampling of our likelihood function using {\tt emcee} \citep{foreman-mackeyEmceeMCMCHammer2013}.  The MCMC is run for a number of steps exceeding 50 times the autocorrelation time to ensure convergence. Running can be slow due to the large number of posterior samples of physical parameters and completeness samples, so we limit the number of posterior sample to 1000. Also, MCMC sampling may struggle to converge with more than 50 bins. Therefore, we divide the period-radius space into multiple patches if more than 50 bins are required, run them separately, and then combine them together. The division is only applied along the orbital period axis because the uncertainties in period are significantly low and can be ignored, so the covariance of occurrence rates is not affected.

\section{Results}
\subsection{Planet candidates}\label{subsec:planet candidates}
From our \textit{TESS}-SPOC FGK main-sequence stellar sample, we first run our detection pipeline for all of them as described in Section~\ref{subsec:detection}. Then, using the \texttt{RAVEN} pipeline (Section~\ref{subsec:vetting&validation}), we identify {1361} planet candidates from our stellar sample, with periods and radii ranging from 0.5 to 16\,days and 2 to {20}\,$R_{\rm \earth}$, respectively. However, the sample is not perfectly clean, particularly for candidates exhibiting orbital period aliases. 

As discussed in Section~\ref{subsec:vetting&validation}, our models are not designed to handle period aliases for planets; an incorrect period would introduce biases into our occurrence estimates. We therefore manually inspect the data by folding the light curve at different aliases of the detected period, excluding only candidates that display clear period aliasing. This yields a final sample of {1301} planet candidates. The number can vary depending on the models and thresholds applied, but the resulting occurrence rates should remain consistent. This consistency arises because our occurrence rate model explicitly accounts for detection completeness and candidate reliability, both of which change alongside different models and thresholds. We further assess the statistical robustness of our occurrence rates in Section~\ref{subsec:thresholds effect} by applying various models and thresholds.

As introduced in Section~\ref{subsubsec:bayesian}, we account for the parameter uncertainties of our planet candidates to achieve a robust occurrence rate estimation. This is accomplished using \texttt{juliet} \citep[][]{espinozaJulietVersatileModelling2019}, which incorporates GP light curve modelling, transit modelling, and nested sampling. Detailed parameter settings and prior distributions are provided in Lafarga et al. (submitted). Both the nested sampling posteriors and priors are stored for our occurrence rate calculations.

\subsection{\textit{TESS} planetary occurrence rates}\label{subsec:TESS occurrence rate}
\begin{figure}
    \centering
    \includegraphics[width=\columnwidth]{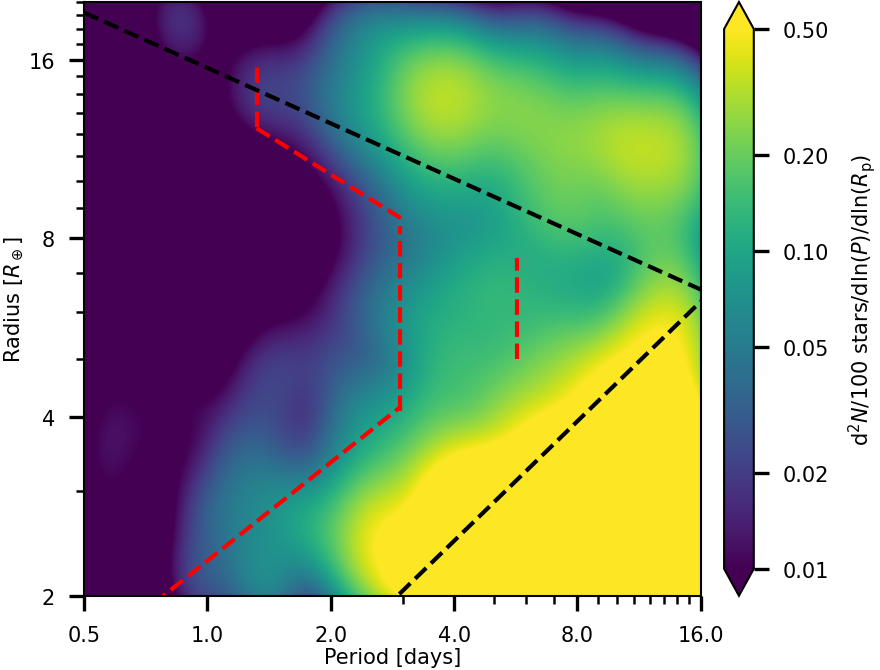}
    \caption{wKDE in period and radius space, with the colour scale normalized logarithmically for better visibility and higher dynamic range. Two Neptunian desert boundaries and a potential ``ridge'' region estimated from \textit{Kepler} are shown with dashed black lines \citep{mazehDearthShortperiodNeptunian2016} and dashed red lines \citep{castro-gonzalezMappingExoNeptunianLandscape2024}. The bandwidth of KDE is {0.3}, which is manually tuned for clearly showing the general features like hot Jupiters and desert boundaries. {The range of our colour scale is limited to match that of Fig.~\ref{fig:occurrence-rate} to have consistent visibility.}}
    \label{fig:wKDE}
\end{figure}

Given the final set of planet candidates, together with our combined completeness maps and reliabilities, we first estimate the occurrence rate distribution via the wKDE method (Section~\ref{subsubsec:IDEM}). Fig.~\ref{fig:wKDE} shows the wKDE result. We can clearly see the differences in occurrence rate density across various regions of the parameter space. {Known features such as the Neptunian desert, together with planet populations like HJs and the abundance of sub‑Neptunes, are all distinctly recovered.} Although the wKDE provides a quick and intuitive view of the occurrence-rate distribution, we still need a more statistically robust model to quantify occurrence rates and compare them with results from the literature.

\begin{figure*}
    \centering
    \includegraphics[width=\textwidth]{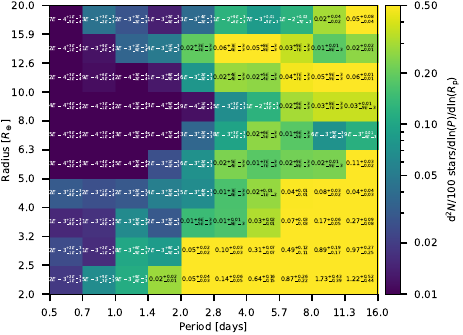}
    \caption{Occurrence rates of planet candidates around \textit{TESS}-SPOC FGK main-sequence stars. Each cell displays the average occurrence as a percentage (number of planets per 100 stars), along with the corresponding uncertainties (the difference between the 16th and 84th percentiles and the median). For clarity, text in cells with an occurrence rate below 0.01\% is displayed in white. The colour of each cell encodes the occurrence rate density, ${\mathrm{d}^2 N_\mathrm{p}}/(\mathrm{d}\ln P\,\mathrm{d}\ln R_\mathrm{p})$, per 100 stars. The logarithmic colour scale is truncated at 0.5 to enhance visibility. We also provide an \href{https://cuikaiming.com/TESS-SPOC-Occurrence-Rate/}{online interactive plot} that allows users to easily select any region and obtain its total occurrence rate.}
    \label{fig:occurrence-rate}
\end{figure*}

Thus, we quantify the occurrence rate within our \textit{TESS} FGK main-sequence stellar sample using the Bayesian method described in Section~\ref{subsubsec:bayesian}. Fig.~\ref{fig:occurrence-rate} presents the occurrence rates on the orbital period-radius plane, using a 10-by-10 grid with both axes in logarithmic scale. The value in each bin represents the average occurrence as a percentage, that is, the number of planets per 100 stars.
Generally, our quantified result shows consistent features with the wKDE result in Fig.~\ref{fig:wKDE}. 
The overall occurrence in our parameter space ($0.5 < P < 20\,{\rm days},\  2 < R_\mathrm{p} < {20}\,R_{\earth}$) is $9.4^{+0.7}_{-0.6}\%$. The direct comparison with Kepler is conducted in Section~\ref{subsec:TESS-Kepler-occ-comp}.

This occurrence map reveals several notable features while providing direct quantitative estimates of occurrence rates for regions such as the distinct Neptunian desert, the piled-up HJs, and the abundance of sub-Neptunes. However, within the scope of this paper, we will focus briefly on HJs {and the Neptunian desert}, comparing our results with previous studies in Section~\ref{subsec:HJ-occ}. Detailed investigations of the those features will be reserved for future work.

The occurrence rate for each bin is provided in the Appendix Table~\ref{tab:TESS-SPOC 10x10 table}. However, we also offer an online interactive plot\footnote{\url{https://cuikaiming.com/TESS-SPOC-Occurrence-Rate/}} where users can click and select any region using the tools like box select or lasso select on the side toolbar to instantly obtain the total occurrence rate and associated uncertainties from the title. We recommend that readers use this tool rather than manually calculating the sum and propagating errors.
We have also made the posterior distributions of occurrence rate density for each cell available online\footnote{\href{https://doi.org/10.5281/zenodo.17804280}{10.5281/zenodo.17804280}}, facilitating future applications and  comparisons.

\subsection{\textit{Gaia} 100\,pc Volume-limited Occurrence Rate}\label{subsec:100pc}
Enabled by \textit{Gaia}'s precise astrometry results and strategically selecting stars within 100\,pc by \textit{TESS}-SPOC, here we tentatively present an analysis of volume-limited planetary occurrence rates for FGK main-sequence stars, which is an important comparison that has rarely been conducted before. Utilizing data from \textit{Gaia} and \textit{TESS}, we achieve this by cross-matching the \textit{Gaia} 100\,pc sample with the TIC catalogue. The \textit{Gaia} Nearby Star Catalogue \citep{smartGaiaEarlyData2021} comprises 331,312 objects within 100\,pc. We perform a cross-match with TIC using an 8 arcsecond radius and selected the closest matches, resulting in 329,921 common objects, a completeness rate of approximately 99.6\%.

\begin{figure}
    \centering
    \includegraphics[width=\columnwidth]{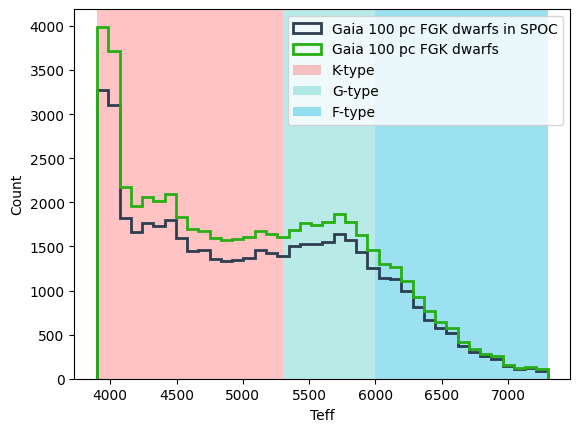}
    \caption{Temperature distributions of FGK main-sequence stars within 100 pc from \textit{Gaia} data, compared to those observed by \textit{TESS}-SPOC across the 1--55th sectors. Shaded regions represent the temperature boundaries for different FGK spectral types.}
    \label{fig:Gaia100pc_teff_comp}
\end{figure}

Subsequently, we identified FGK main-sequence stars (with effective temperatures between 3900--7300\,K and $\log g > 3.5$) based on TIC parameters, yielding a subset of 56,197 stars. Of these, \textit{TESS}-SPOC observed 48,622 within Sectors 1--55. We also evaluated the completeness of our cross-matched sample across different temperature ranges as illustrated in Fig.~\ref{fig:Gaia100pc_teff_comp}. Although the number of \textit{TESS}-SPOC cross-matched samples is relatively lower than that of \textit{Gaia}'s FGK dwarfs within the same distance range (100\,pc), quantitatively, the completeness for F, G, and K types all remain above 85\%. 

\begin{figure}
    \centering
    \includegraphics[width=\columnwidth]{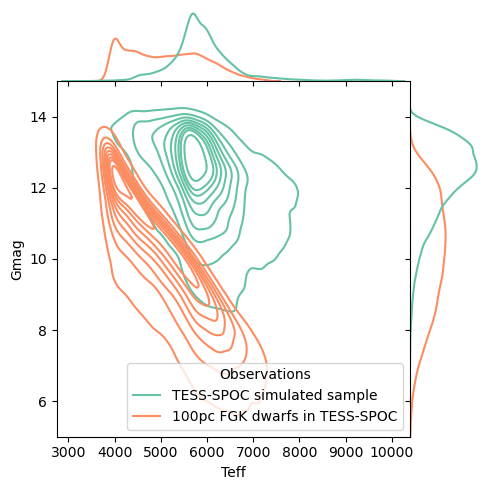}
    \caption{2D KDE contours showing the effective temperature and Gaia G-band magnitude distributions for our \textit{TESS}-SPOC simulated sample and for Gaia 100\,pc FKG main-sequence stars observed by \textit{TESS}-SPOC.}
    \label{fig:teff-Gmag}
\end{figure}

We then apply the same pipeline to the observed volume-limited sample, using the non-uniform FP models, setting the NSFP, FPs, and SNR thresholds to 0.9, 0.9, and 10, respectively. This yields {125} planetary candidates within our parameter space range ($0.5 < P < 16\,{\rm days},\  2 < R_\mathrm{p} < {20}\,R_{\earth}$). We further fit their planetary parameters and estimate the occurrence rate by incorporating the completeness map and using the Bayesian method as previously described. Notably, because of the significant differences in stellar properties between magnitude-limited and volume-limited samples, all completeness maps must be recalculated. The 100\,pc sample, in particular, contains a higher proportion of K dwarfs, and its stars tend to be apparently brighter. Fig.~\ref{fig:teff-Gmag} shows effective temperature and Gaia Gmag distributions of the observed 100\,pc sample and our simulated sample.

\begin{figure}
        \centering
        \includegraphics[width=\columnwidth]{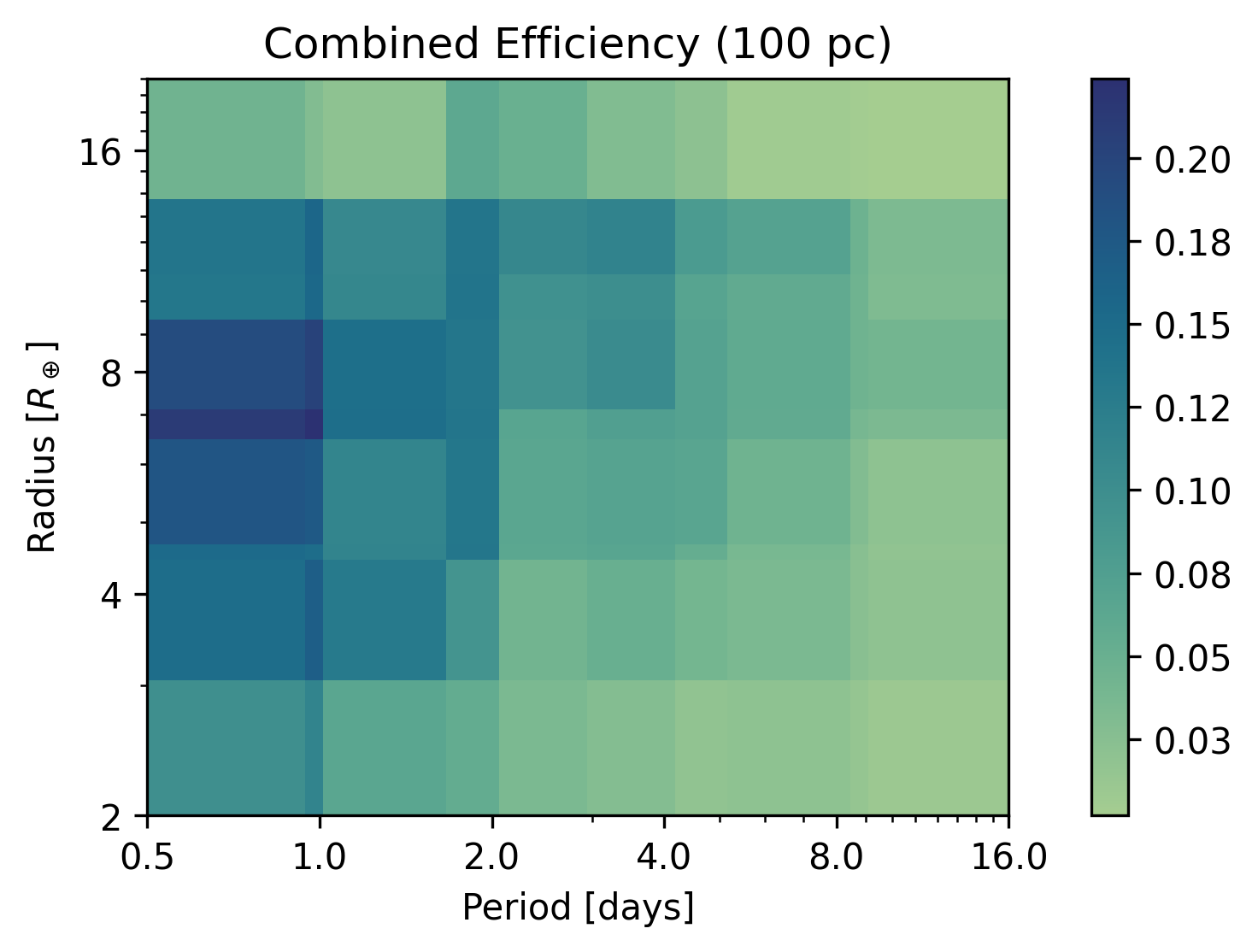}
        \caption{{Combined completeness} of \textit{Gaia} 100\, pc FGK main-sequence stars.}
        \label{fig:comp-100pc}
\end{figure}

To address this, importance sampling is applied to our simulated sample to match the effective temperature and Gmag distributions of the 100\,pc sample: the simulated sample is weighted by the ratio between the target (100\,pc sample) probability density and our simulated probability density. Effective temperature and Gmag are used because, for main-sequence stars, most other stellar properties, such as stellar radius, light curve noise levels, and crowdings are either closely related to these two parameters or can be assumed to have the same distribution in the simulated and real samples. Note that there are a few faint stars (Gmag > 14) in the 100\,pc sample that are not included in our TESS-SPOC simulated sample, so their contributions to the completeness may not be accurately modelled.
Fig.~\ref{fig:comp-100pc} shows the {combined} completeness map for the 100\,pc sample. We reduce the number of bins to mitigate fluctuations caused by a few high-weight points.

\begin{figure}
    \centering
    \includegraphics[width=\columnwidth]{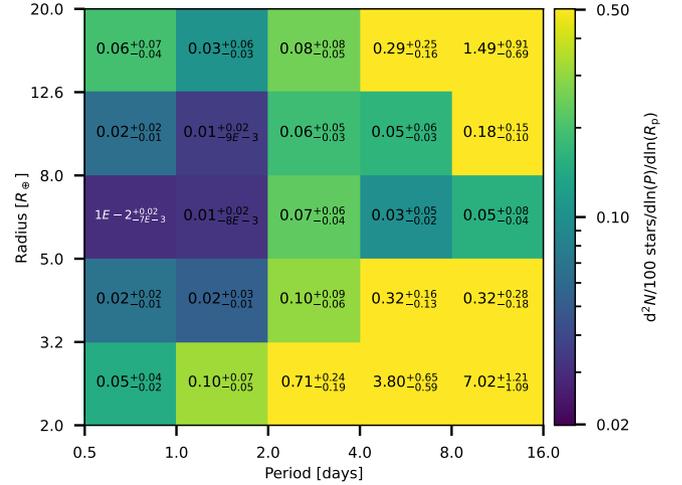}
    \caption{Similar to Fig.~\ref{fig:occurrence-rate}, occurrence rates of \textit{Gaia} 100\,pc sample.}
    \label{fig:100pc-occurrence}
\end{figure}

\begin{figure}
    \centering
    \includegraphics[width=\columnwidth]{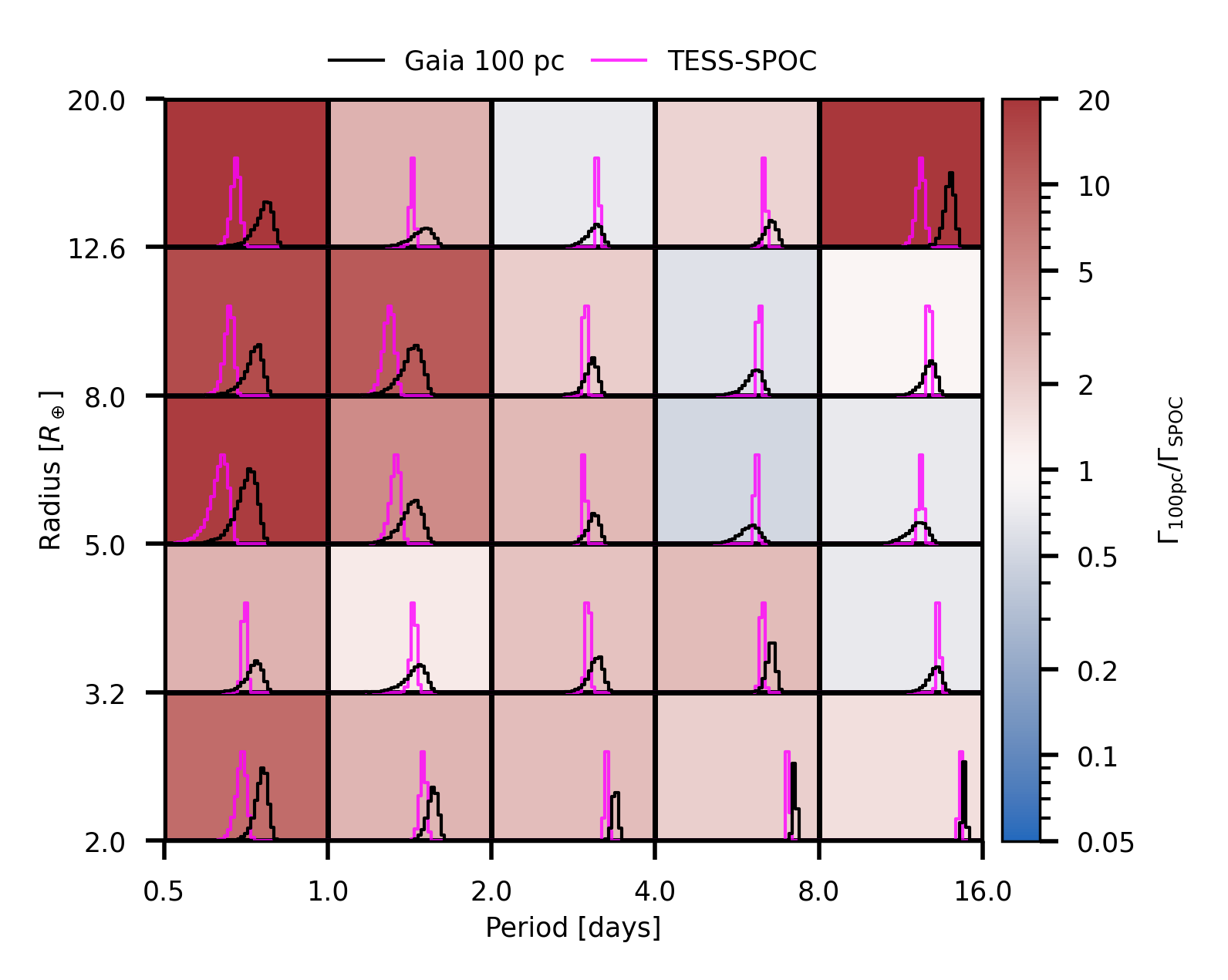}
    \caption{Posterior distributions of planetary occurrence rates for both the \textit{Gaia} 100\,pc sample and our \textit{TESS}-SPOC sample. The black and fuchsia lines represent the \textit{Gaia} 100\,pc and \textit{TESS}-SPOC samples, respectively. In each cell, the x-axis of occurrence rates ranging from $1\times10^{-6}\%$ to 50\% with logarithmic scale, and the heights of the distributions are scaled according to the highest histogram in each cell. The background diverging colour map indicates the ratio of occurrence rates between the two samples. The colour scale is manually centred at one; the actual minimum ratio is approximately 0.4.}
    \label{fig:100pc-vs-SPOC}
\end{figure}

Following the same procedure, we estimate the occurrence rate for the 100\,pc sample. However, due to importance sampling, the numbers for injection and recovery are not feasible to obtain. Therefore, we use the expected values of completenesses directly, without modelling them with Beta distributions. Fig.~\ref{fig:100pc-occurrence} presents the occurrence rate distributions, and the overall occurrence rate in our parameter space is $15.4^{+1.6}_{-1.5}\%$, which is {higher} than our \textit{TESS}-SPOC result. In detail, Fig.~\ref{fig:100pc-vs-SPOC} compares the posteriors of our magnitude-limited and volume-limited samples within each bin. The colour map in Fig.~\ref{fig:100pc-vs-SPOC} is scaled so that the white midpoint represents where the \textit{Gaia} 100\,pc and \textit{TESS}-SPOC median occurrence rates are equal. On this scale, bluish colours indicate that the \textit{TESS}-SPOC occurrence rate is higher than that of \textit{Gaia} 100\,pc, whereas reddish colours indicate that the \textit{Gaia} 100\,pc occurrence rate is higher than that of \textit{TESS}-SPOC. 

As expected, the \textit{Gaia} 100\,pc occurrence rate shows larger uncertainties because of limited sample size. In most bins, the {\textit{Gaia} 100\,pc sample generally shows higher occurrence rates, although these rates are consistent within $1\sigma$. This elevation can be attributed to the sample's larger proportion of K-type stars, reflecting the known temperature dependence of planetary occurrence rates \citep[e.g.,][]{kunimotoSearchingEntiretyKepler2020}. The especially high occurrence rates in the 100\,pc sample for shorter periods (< $1\sim2$\,days) are due to the very small number of detections and the limited size of the stellar sample, which is much smaller than that of the \textit{TESS}-SPOC sample. Because a smaller stellar sample sets a relatively high lower limit on the minimum measurable occurrence rate, even a few detected planets can result in a noticeably higher inferred rate compared to the \textit{TESS}-SPOC sample, which contains far more stars.}


\subsection{Hot Jupiters and the Neptunian desert}\label{subsec:HJ-occ}
As one of the most prominent features in planet occurrence rate distributions, the occurrence of hot Jupiters has been widely studied. 
\citet{howardPLANETOCCURRENCE0252012} found $0.4\pm0.1\%$ for \textit{Kepler} GK-type stars. Later, \citet{fressinFALSEPOSITIVERATE2013} reported $0.43\pm0.05\%$ for \textit{Kepler} FGK dwarfs, and 
{\citet{santerneSOPHIEVelocimetryKepler2016} obtained $0.47\pm0.08\%$ for Kepler F5--K5 V-type stars. }
\citet{petiguraCaliforniaKeplerSurveyIV2018} estimated $0.57^{+0.14}_{-0.12}\%$ for \textit{Kepler} FGK-type stars. \citet{zhouTwoNewHATNet2019} measured an HJ occurrence rate of $0.41\pm0.10\%$ for early \textit{TESS} AFG-type stars; subsequently, \citet{beleznayExploringDependenceHot2022} derived $0.33\pm0.04\%$ for AFG-type stars in a larger sample. Studies of M dwarfs \citep[e.g.,][]{ganOccurrenceRateHot2023,bryantOccurrenceRateGiant2023} indicate lower occurrence rates toward lower stellar masses. Others have examined HJ occurrence as a function of metallicity, age, and evolutionary stage \citep[e.g.,][]{guoMetallicityDistributionHot2017,osbornInvestigatingPlanetMetallicity2020,yeePeriodDistributionHot2023,miyazakiEvidenceThatOccurrence2023,temminkOccurrenceRateHot2023}.

We refine our HJ parameter space to $0.8 < P < 10\,{\rm days}$ and $8 < R_{\rm p} < 20\,{\rm R_{\earth}}$, identifying {432} HJ candidates in our \textit{TESS}-SPOC sample. This number is broadly consistent with the predictions of \citet{yeeHowCompleteAre2021}. Using the same method, we derive an overall HJ occurrence rate of $0.39^{+0.03}_{-0.02}\%$. This value is {fully consistent with} previously reported results for the \textit{Kepler} FGK and some \textit{TESS} AFG samples. 
In our analysis of the \textit{Gaia} 100\,pc sample, we also estimate an overall HJ occurrence rate of $0.42^{+0.16}_{-0.12}\%$. This value with its relative larger uncertainty is in agreement with the above estimations within $1\sigma$. 

{Due to the low number of planets in the Neptunian desert, as far as we are aware, no previous work has directly reported its overall occurrence rate. However, we can estimate the Neptunian desert occurrence rate in the \textit{Kepler} data using our posterior distribution derived from \citet{hsuOccurrenceRatesPlanets2019}. Adopting the desert boundary defined by \citet{castro-gonzalezMappingExoNeptunianLandscape2024} and assuming that the occurrence rate within each bin is uniformly distributed on a logarithmic scale, we can calculate the overall Neptunian desert occurrence rate from Kepler data. Since much of the desert region remains undetected, we adopt the 95\textsuperscript{th} percentile as a conservative upper limit, which is $0.69\%$. Within the boundary, our sample has 108 candidates, and applying the same desert boundary to our posterior distributions yields an overall occurrence rate for \textit{TESS}-SPOC of $0.08\pm0.01\%$, with a 95\textsuperscript{th} percentile upper limit of $0.10\%$. Our results provide the first direct constraint on the overall occurrence rate of the Neptunian desert within these boundaries. Our results are consistent with the \textit{Kepler} upper limit, in contrast with the potentially higher desert occurrence rate in \textit{TESS} data reported in the preliminary results of \citet{meltonDIAmanteTESSAutoRegressive2024b}.}

\section{Discussion}\label{sec:discussion}

\subsection{Effect of different models and thresholds}\label{subsec:thresholds effect}
Our occurrence rate calculations rely heavily on our vetting and validation models; they affect both the number of detected planets and the completeness estimates. Nevertheless, the occurrence rate estimation process should yield robust results under the following ideal conditions: (1) the stellar properties in our simulations match those of the observed stellar sample; (2) the simulated sample is processed by the same pipeline as the observed sample, with no data leakage; and (3) the vetting and validation models produce statistically reliable posterior probabilities.

To meet these requirements: (1) we select the simulated sample from the same sources as the observed stellar sample, removing only binaries and known planets; (2) we evaluate completeness with the same pipeline used for detection, maintain a clear training/test split, and never use the observed light curves when training and evaluating models (except for NSFP, which are sampled from observed BLS peaks, but their number is limited); and (3) although it is impossible to know the true posterior probabilities of candidates, except for confirmed planets and FPs, our machine learning models are well calibrated, supporting the reliability of their probabilities. In addition, adopting a relatively high decision threshold helps ensure that the selected candidates can be treated as confirmed planets.

\begin{figure*}
    \centering
    \includegraphics[width=0.9\textwidth]{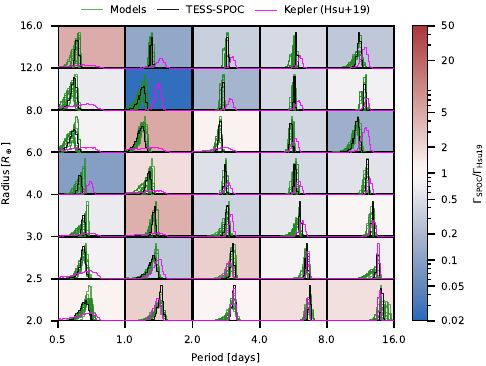}
    \caption{Similar to Fig.~\ref{fig:100pc-vs-SPOC}, this figure shows the posterior distributions of planetary occurrence rates for our \textit{TESS}-SPOC sample using various thresholds, alongside \textit{Kepler} results. The fuchsia lines represent the \textit{Kepler} findings from \citet{hsuOccurrenceRatesPlanets2019}, while the green curves correspond to our model's results across 18 different threshold combinations. The black line denotes our selected model, which uses a non-uniform model with NSFP and FP thresholds of 0.9 and an SNR threshold of 10. In each cell, the x-axis of occurrence rates ranging from $1\times10^{-5}$\% to 100\% with logarithmic scale. The background diverging colour map shows the ratio of occurrence rates between our chosen model and the \textit{Kepler} results. The colour map scale is manually centred at one; the actual maximum ratio is approximately five.}
    \label{fig:post-comp}
\end{figure*}

Based on the discussion above, we test the robustness of our occurrence rate estimates by varying the models and thresholds. Specifically, we explore two validation models (uniform and non-uniform), FP thresholds of 0.90, 0.95, and 0.99, and SNR thresholds of 10, 15, and 20, resulting in 18 distinct combinations. For each of these 18 cases, we estimate the occurrence rate using the same hierarchical Bayesian procedure and obtain their corresponding posterior distributions. For convenience, we use the same bin configuration as \citet{hsuOccurrenceRatesPlanets2019}, since we will be comparing our posteriors with theirs as well (see Section~\ref{subsec:TESS-Kepler-occ-comp}). Figure~\ref{fig:post-comp} shows the posterior distributions of the occurrence rates for all combinations considered, highlighting both our selected threshold and the results of \citet{hsuOccurrenceRatesPlanets2019}.

We measure the relative scattering of occurrence rate posterior distributions by calculating the median absolute deviation (MAD) of median occurrence rates over the median uncertainty for posteriors in each cell. Overall, our posterior distributions remain self-consistent across most threshold combinations in most bins, only {four} bins show the MAD over median uncertainties greater than one. These cases can be found in Fig.~\ref{fig:post-comp} by their increased scatter and always occur at the {edges} of the parameter space. We highlight the most pronounced discrepancies in Appendix~\ref{appendix}, and we show that the main differences arise from using excessively high FP threshold or uniform models. 

In principle, as discussed, the occurrence rate should remain consistent regardless of the FP or SNR threshold and model used. However, at higher thresholds, estimating completeness becomes difficult because the recovered simulated sample size in some bins is very small, sometimes zero, leading to highly uncertain estimates. A similar issue arises for the uniform models: assuming planets occur uniformly in orbital period and planetary radius space, the simulated uniform sample is dominated by observational bias, yielding very few observable planets at longer periods and smaller radii. Consequently, the simulated sample used to estimate FP completeness can be extremely limited, sometimes nearly zero. Thus it suffers from both too few detections and too few injections. This is evident in Fig.~\ref{fig:completeness-grid-uniform}, where the bottom-right bin shows zero completeness. 

Given these results and analyses, we can confidently state that our occurrence rate estimates are robust.

\subsection{Effect of RUWE}\label{subsec:RUWE-effect}
{As a commonly used binarity metric, RUWE may indicate different planetary occurrence rates between low and high values.
To quantify differences in occurrence rates between low RUWE (<1.05) and high RUWE (>1.05) stars, we divide our stellar sample into these two groups, resulting in 54\% low RUWE stars and 46\% high RUWE stars. In our {\tt RAVEN} pipeline, RUWE affects our HTP and HEB priors: when RUWE is less than 1.05, we reduce the HTP and HEB priors by preset factors, and when RUWE exceeds 1.4, we increase the HTP and HEB priors by specific factors \citep[detailed in][]{hadjigeorghiouRAVENRAnkingValidation2025}.
Thus, we calculate completeness by applying the low RUWE prior to low RUWE cases and the high RUWE prior (>1.05) to high RUWE cases. Next, we separate our high-confidence planet candidates into low RUWE and high RUWE groups and apply the same occurrence rate estimation methodology to each. This allows us to compare their occurrence rates in detail.}

\begin{figure}
    \centering
    \includegraphics[width=\columnwidth]{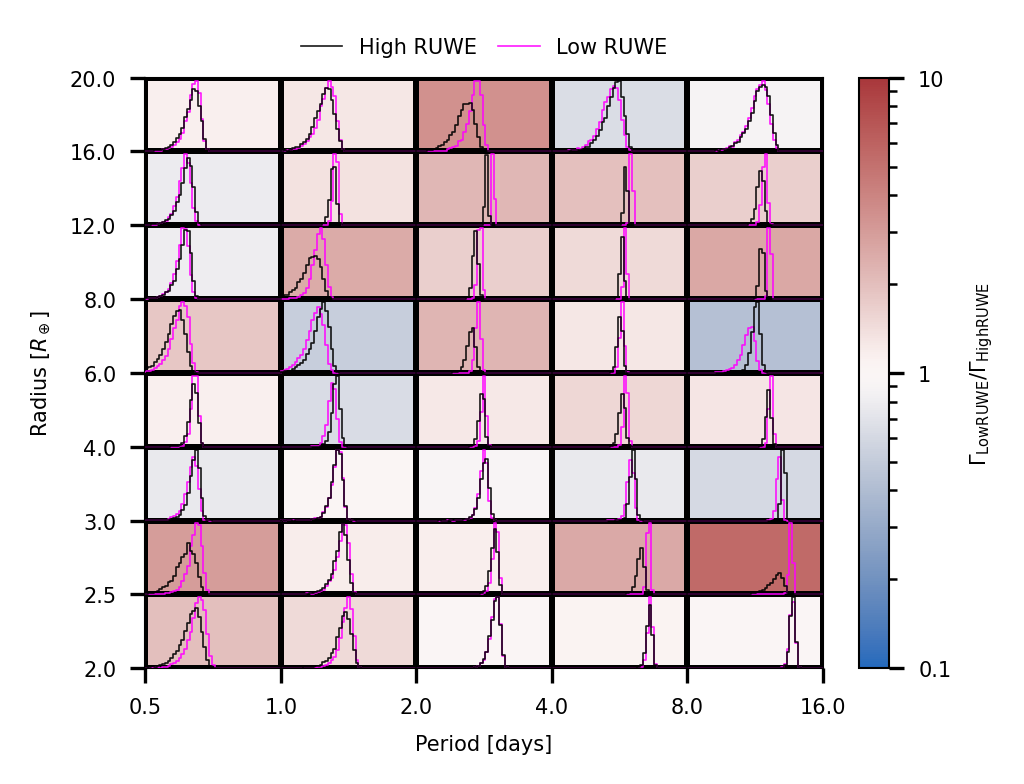}
    \caption{{Similar to Fig.~\ref{fig:100pc-vs-SPOC}, this figure shows the posterior distributions of planetary occurrence rates for low RUWE ($<1.05$) and high RUWE ($>1.05$) stars. The black lines indicate the high RUWE sample and the fuchsia lines indicate the low RUWE sample. The background diverging colour maps the ratio of median occurrence rates between low RUWE and high RUWE sample.}}
    \label{fig:ruwe-analysis}
\end{figure}

{Figure~\ref{fig:ruwe-analysis} shows a comparison of planetary occurrence rates between stars with low and high RUWE values. Even though all their differences are within one magnitude, the high RUWE sample generally exhibits lower occurrence rates. This trend may be related to the prevalence of unresolved binaries, which can both influence RUWE measurements and impact planet detection. Our results suggest that unresolved binaries may suppress planet formation or reduce the likelihood of detecting planets in these systems. Further investigation into the relationship between RUWE, binarity, and planetary systems will require a deeper understanding of the RUWE parameter, which is expected to improve with the release of Gaia DR4.}

\subsection{\textit{TESS} and \textit{Kepler} occurrence rate comparison}\label{subsec:TESS-Kepler-occ-comp}
There are many occurrence rate estimates based on \textit{Kepler} FGK stars \citep[e.g.,][]{kunimotoSearchingEntiretyKepler2020,hsuOccurrenceRatesPlanets2019}. Here, we compare our results with those of \citet{hsuOccurrenceRatesPlanets2019} within the same parameter space range ($0.5 < P < 16\,{\rm days},\  2 < R_\mathrm{p} < 16\,R_{\earth}$). Since \citet{hsuOccurrenceRatesPlanets2019} only provides occurrence rates with uncertainties at the 16th and 84th percentiles, we fit these uncertainty percentiles with Gamma distributions to directly compare their occurrence rate posteriors with ours. For occurrences where \citet{hsuOccurrenceRatesPlanets2019} only provides upper limits due to zero detections, we assume the mean of a Gamma distribution is half of the 84th percentile value, which roughly matches their posterior distributions shown in Fig.~5 of their paper. 

For the overall occurrence rate, \citet{hsuOccurrenceRatesPlanets2019} gives $11.4^{+1.2}_{-1.1}\%$ (this value is estimated from the above posteriors; the direct sum over bins with non-zero detections yields $11.0\%$. Since we do not have access to their posterior samples, small differences may be present). Our \textit{TESS}-SPOC occurrence rate estimation within $2 < R_\mathrm{p} < 16\,R_{\earth}$ is $9.2^{+0.7}_{-0.6}\%$, consistent within $1.5\sigma$. 
{We also adopt the largest common parameter range from \citet[][$0.75 < P < 12.5\,{\rm days},\ 2 < R_\mathrm{p} < 16\,R_{\earth}$]{kunimotoSearchingEntiretyKepler2020}. Compared to their overall median value of 6.7\% (not accounting for the upper limit in the desert region), our result of $7.2 \pm 0.5\%$ is consistent within $1\sigma$.}

To have a detailed comparison, we use the same bin configuration as \citet{hsuOccurrenceRatesPlanets2019} and run our Bayesian model as described in Section~\ref{subsec:TESS occurrence rate}. Fig.~\ref{fig:post-comp} compares the posterior distributions of occurrence rate estimates for each bin. The colour scale of this plot is set to be one when the occurrence of \textit{TESS}-SPOC and \textit{Kepler} are equal. Thus the bluish bins show our occurrence rates are lower than \textit{Kepler}, while the reddish colour indicate our \textit{TESS}-SPOC occurrence rate is higher than \textit{Kepler}. 

From the Fig.~\ref{fig:post-comp}, we can see that our occurrence rate posterior distributions are significantly tighter than those of \citet{hsuOccurrenceRatesPlanets2019} in most bins. On average, the $1\sigma$ uncertainty ranges of our posteriors (defined as the difference between the 84\% and 16\% percentiles) are roughly an order of magnitude smaller than those reported by \citet{hsuOccurrenceRatesPlanets2019}. This improvement is particularly evident near the Neptunian desert region. Within the desert, our posteriors provide better constraints, reducing some of the uncertainties present in that region. For example, in the bins with $0.5<P<1.0\,{\rm days}$, $4.0 < R_\mathrm{p} < 6.0\,R_{\earth}$ and $1.0<P<2.0\,{\rm days}$, $8.0 < R_\mathrm{p} < 12.0\,R_{\earth}$, \citet{hsuOccurrenceRatesPlanets2019} report anomalously high occurrence rates compared to their surrounding bins, whereas our results show lower values that are more consistent with those of the adjacent desert regions, resulting in a more well-defined Neptunian desert.

\subsection{Future Prospects}\label{subsec:future}
Given our comprehensive pipeline and occurrence rate results, numerous directions remain to be explored. For instance, future studies can investigate occurrence rates across various stellar types, metallicities, and evolutionary stages, with particular attention to sub-Neptunes and hot Jupiters. Additionally, it will be valuable to precisely quantify the boundaries of the Neptunian desert, as well as determine the locations and significance of the ridge and savanna \citep[][]{bourrierDREAMOrbitalArchitecture2023,castro-gonzalezMappingExoNeptunianLandscape2024}, and to compare these findings with theoretical models of planet formation and evolution. We plan to address these topics in future research.

Additionally, extending our pipeline to more \textit{TESS} FFI observations, specifically to sectors beyond sector 55, is a straightforward way to expand the sample size and obtain higher precision.

\section{Conclusion}
In this study, we measure the occurrence rates of \textit{TESS} close-in exoplanets orbiting FGK main sequence stars. Our main results are summarized as below:
\begin{itemize}
    \item We measured the occurrence of close-in exoplanets around FGK main-sequence stars using four years of \textit{TESS}-SPOC observations. Our analysis spans orbital periods of 0.5--16 days and radii of 2--{20}$\,R_{\earth}$, evaluated on a $10 \times 10$ grid in period-radius space. The results are robust, as confirmed by our tests.
    \item For the magnitude-limited \textit{TESS}-SPOC sample, the overall occurrence rate is {$9.4^{+0.7}_{-0.6}\%$, and the distributions are consistent with the Kepler-based estimates \citep[][]{hsuOccurrenceRatesPlanets2019} across most of the commonly studied parameter space.}
    \item The \textit{Gaia} Nearby Star Catalogue, volume-limited subsample within 100\,pc exhibits a similar structure in the occurrence map but systematically higher rates with larger uncertainties, yielding an overall occurrence of $15.4^{+1.6}_{-1.5}\%$.
    \item In the HJ regime, we find an occurrence rate of $0.39^{+0.03}_{-0.02}\%$ in the full \textit{TESS}-SPOC sample. Our results are consistent with Kepler FGK results within $1\sigma$. These findings are also consistent with those from the 100 pc sample, which has an HJ occurrence rate density of $0.42^{+0.16}_{-0.12}\%$.
    \item {For the Neptunian desert, adopting the boundary from \citet{castro-gonzalezMappingExoNeptunianLandscape2024}, we find an overall occurrence rate of $0.08\pm0.01\%$, with a 95\textsuperscript{th} percentile upper limit of $0.10\%$. Our results significantly enhance the precision of occurrence rate estimates made possible by \textit{TESS}.}
    \item Together, these comparisons demonstrate that our systematic detection, validation, and occurrence rate estimation reliably provide stronger constraints on the demographics of close-in exoplanets orbiting FGK main-sequence stars.
\end{itemize}

\textit{TESS}-SPOC provides a powerful platform for exoplanet statistics. Extending to additional sectors, broader period-radius ranges, and higher-dimensional dependencies (e.g., stellar mass and metallicity), alongside continued refinement of our completeness and reliability models, will sharpen the empirical constraints and test theoretical predictions.

\section*{Acknowledgements}
This paper includes data collected by the \textit{TESS} mission. Funding for the \textit{TESS} mission is provided by the NASA Explorer Program. Resources supporting this work were provided by the NASA High-End Computing (HEC) Program through the NASA Advanced Supercomputing (NAS) Division at Ames Research Center for the production of the SPOC data products. This research was funded by the UKRI (Grants ST/X001121/1, EP/X027562/1). Some of the language has been polished with the assistance of OpenAI API Service \citep{openaiGPT4oModel2024}.

\section*{Data Availability}
The \textit{TESS}-SPOC data used in this paper can be found in Mikulski Archive for Space Telescopes \citep{caldwellTESSLightCurves2020}.
Our \textit{TESS}-SPOC occurrence rate posterior sample is available \citep[][]{cuiSamplePosteriorDistributions2025}.



\bibliographystyle{mnras}
\bibliography{Neptune} 




\appendix
\section{Occurrence rate tables}
Table~\ref{tab:TESS-SPOC 10x10 table} lists the occurrence rate values and uncertainties of Fig.~\ref{fig:occurrence-rate}. Table~\ref{tab:TESS-SPOC 100 pc table} lists the occurrence rate values and uncertainties of Fig.~\ref{fig:100pc-occurrence}.

\begin{table*}
\caption{Occurrence rates (number of planets per 100 stars) of our 10-by-10 bins of Fig.~\ref{fig:occurrence-rate}. Values are reported to two significant figures.}
\label{tab:TESS-SPOC 10x10 table}
\renewcommand{\arraystretch}{1.28}
\begin{tabular}{ccc @{\ \vline\ } ccc}
\hline
Period ranges (days) & Radius ranges ($R_{\earth}$) & Occurrence rates (\%)
   & Period ranges (days) & Radius ranges ($R_{\earth}$) & Occurrence rates (\%) \\ 
\hline
0.50 -- 0.71 & 2.00 -- 2.52 & $0.0015 [+0.0024, -0.0011]$ &
0.50 -- 0.71 & 2.52 -- 3.17 & $0.0019 [+0.0017, -0.0012]$ \\
0.50 -- 0.71 & 3.17 -- 3.99 & $0.0012 [+0.0011, -0.00079]$ &
0.50 -- 0.71 & 3.99 -- 5.02 & $0.0024 [+0.0010, -0.00084]$ \\
0.50 -- 0.71 & 5.02 -- 6.32 & $0.00033 [+0.00050, -0.00024]$ &
0.50 -- 0.71 & 6.32 -- 7.96 & $0.00047 [+0.00048, -0.00031]$ \\
0.50 -- 0.71 & 7.96 -- 10.02 & $0.00043 [+0.00041, -0.00027]$ &
0.50 -- 0.71 & 10.02 -- 12.62 & $0.00021 [+0.00031, -0.00016]$ \\
0.50 -- 0.71 & 12.62 -- 15.89 & $0.00018 [+0.00029, -0.00013]$ &
0.50 -- 0.71 & 15.89 -- 20.00 & $0.00066 [+0.0011, -0.00049]$ \\
0.71 -- 1.00 & 2.00 -- 2.52 & $0.0053 [+0.0049, -0.0033]$ &
0.71 -- 1.00 & 2.52 -- 3.17 & $0.0013 [+0.0019, -0.00097]$ \\
0.71 -- 1.00 & 3.17 -- 3.99 & $0.0014 [+0.0015, -0.00092]$ &
0.71 -- 1.00 & 3.99 -- 5.02 & $0.0021 [+0.0011, -0.00085]$ \\
0.71 -- 1.00 & 5.02 -- 6.32 & $0.00025 [+0.00043, -0.00019]$ &
0.71 -- 1.00 & 6.32 -- 7.96 & $0.00019 [+0.00030, -0.00014]$ \\
0.71 -- 1.00 & 7.96 -- 10.02 & $0.00017 [+0.00027, -0.00013]$ &
0.71 -- 1.00 & 10.02 -- 12.62 & $0.00091 [+0.00063, -0.00048]$ \\
0.71 -- 1.00 & 12.62 -- 15.89 & $0.00097 [+0.00087, -0.00058]$ &
0.71 -- 1.00 & 15.89 -- 20.00 & $0.0035 [+0.0030, -0.0021]$ \\
1.00 -- 1.41 & 2.00 -- 2.52 & $0.0089 [+0.0090, -0.0059]$ &
1.00 -- 1.41 & 2.52 -- 3.17 & $0.0082 [+0.0062, -0.0049]$ \\
1.00 -- 1.41 & 3.17 -- 3.99 & $0.0051 [+0.0033, -0.0025]$ &
1.00 -- 1.41 & 3.99 -- 5.02 & $0.0023 [+0.0015, -0.0012]$ \\
1.00 -- 1.41 & 5.02 -- 6.32 & $0.00052 [+0.00071, -0.00039]$ &
1.00 -- 1.41 & 6.32 -- 7.96 & $0.00028 [+0.00045, -0.00021]$ \\
1.00 -- 1.41 & 7.96 -- 10.02 & $0.00019 [+0.00031, -0.00014]$ &
1.00 -- 1.41 & 10.02 -- 12.62 & $0.00021 [+0.00033, -0.00015]$ \\
1.00 -- 1.41 & 12.62 -- 15.89 & $0.0028 [+0.0011, -0.00085]$ &
1.00 -- 1.41 & 15.89 -- 20.00 & $0.0022 [+0.0026, -0.0015]$ \\
1.41 -- 2.00 & 2.00 -- 2.52 & $0.022 [+0.017, -0.012]$ &
1.41 -- 2.00 & 2.52 -- 3.17 & $0.0067 [+0.0076, -0.0046]$ \\
1.41 -- 2.00 & 3.17 -- 3.99 & $0.0023 [+0.0026, -0.0016]$ &
1.41 -- 2.00 & 3.99 -- 5.02 & $0.0043 [+0.0026, -0.0020]$ \\
1.41 -- 2.00 & 5.02 -- 6.32 & $0.0021 [+0.0014, -0.0010]$ &
1.41 -- 2.00 & 6.32 -- 7.96 & $0.00047 [+0.00071, -0.00035]$ \\
1.41 -- 2.00 & 7.96 -- 10.02 & $0.00031 [+0.00050, -0.00023]$ &
1.41 -- 2.00 & 10.02 -- 12.62 & $0.0010 [+0.00076, -0.00053]$ \\
1.41 -- 2.00 & 12.62 -- 15.89 & $0.0039 [+0.0013, -0.0011]$ &
1.41 -- 2.00 & 15.89 -- 20.00 & $0.0011 [+0.0018, -0.00085]$ \\
2.00 -- 2.83 & 2.00 -- 2.52 & $0.049 [+0.036, -0.026]$ &
2.00 -- 2.83 & 2.52 -- 3.17 & $0.045 [+0.017, -0.015]$ \\
2.00 -- 2.83 & 3.17 -- 3.99 & $0.011 [+0.0065, -0.0052]$ &
2.00 -- 2.83 & 3.99 -- 5.02 & $0.0053 [+0.0037, -0.0028]$ \\
2.00 -- 2.83 & 5.02 -- 6.32 & $0.0057 [+0.0024, -0.0020]$ &
2.00 -- 2.83 & 6.32 -- 7.96 & $0.0029 [+0.0017, -0.0013]$ \\
2.00 -- 2.83 & 7.96 -- 10.02 & $0.00094 [+0.0010, -0.00063]$ &
2.00 -- 2.83 & 10.02 -- 12.62 & $0.0063 [+0.0021, -0.0018]$ \\
2.00 -- 2.83 & 12.62 -- 15.89 & $0.015 [+0.0030, -0.0027]$ &
2.00 -- 2.83 & 15.89 -- 20.00 & $0.0031 [+0.0035, -0.0022]$ \\
2.83 -- 4.00 & 2.00 -- 2.52 & $0.14 [+0.063, -0.052]$ &
2.83 -- 4.00 & 2.52 -- 3.17 & $0.10 [+0.031, -0.029]$ \\
2.83 -- 4.00 & 3.17 -- 3.99 & $0.013 [+0.011, -0.0080]$ &
2.83 -- 4.00 & 3.99 -- 5.02 & $0.013 [+0.0071, -0.0057]$ \\
2.83 -- 4.00 & 5.02 -- 6.32 & $0.018 [+0.0055, -0.0045]$ &
2.83 -- 4.00 & 6.32 -- 7.96 & $0.0077 [+0.0031, -0.0025]$ \\
2.83 -- 4.00 & 7.96 -- 10.02 & $0.0054 [+0.0026, -0.0021]$ &
2.83 -- 4.00 & 10.02 -- 12.62 & $0.022 [+0.0044, -0.0040]$ \\
2.83 -- 4.00 & 12.62 -- 15.89 & $0.061 [+0.0068, -0.0064]$ &
2.83 -- 4.00 & 15.89 -- 20.00 & $0.0098 [+0.0091, -0.0062]$ \\
4.00 -- 5.66 & 2.00 -- 2.52 & $0.64 [+0.16, -0.15]$ &
4.00 -- 5.66 & 2.52 -- 3.17 & $0.31 [+0.075, -0.065]$ \\
4.00 -- 5.66 & 3.17 -- 3.99 & $0.025 [+0.017, -0.014]$ &
4.00 -- 5.66 & 3.99 -- 5.02 & $0.021 [+0.011, -0.0096]$ \\
4.00 -- 5.66 & 5.02 -- 6.32 & $0.014 [+0.0071, -0.0065]$ &
4.00 -- 5.66 & 6.32 -- 7.96 & $0.020 [+0.0056, -0.0051]$ \\
4.00 -- 5.66 & 7.96 -- 10.02 & $0.0098 [+0.0038, -0.0031]$ &
4.00 -- 5.66 & 10.02 -- 12.62 & $0.020 [+0.0049, -0.0044]$ \\
4.00 -- 5.66 & 12.62 -- 15.89 & $0.054 [+0.0081, -0.0076]$ &
4.00 -- 5.66 & 15.89 -- 20.00 & $0.0067 [+0.010, -0.0050]$ \\
5.66 -- 8.00 & 2.00 -- 2.52 & $0.87 [+0.26, -0.22]$ &
5.66 -- 8.00 & 2.52 -- 3.17 & $0.49 [+0.12, -0.11]$ \\
5.66 -- 8.00 & 3.17 -- 3.99 & $0.071 [+0.032, -0.029]$ &
5.66 -- 8.00 & 3.99 -- 5.02 & $0.039 [+0.015, -0.013]$ \\
5.66 -- 8.00 & 5.02 -- 6.32 & $0.017 [+0.0089, -0.0075]$ &
5.66 -- 8.00 & 6.32 -- 7.96 & $0.014 [+0.0059, -0.0051]$ \\
5.66 -- 8.00 & 7.96 -- 10.02 & $0.024 [+0.0061, -0.0056]$ &
5.66 -- 8.00 & 10.02 -- 12.62 & $0.037 [+0.0072, -0.0066]$ \\
5.66 -- 8.00 & 12.62 -- 15.89 & $0.028 [+0.0086, -0.0074]$ &
5.66 -- 8.00 & 15.89 -- 20.00 & $0.0099 [+0.016, -0.0073]$ \\
8.00 -- 11.31 & 2.00 -- 2.52 & $1.7 [+0.43, -0.39]$ &
8.00 -- 11.31 & 2.52 -- 3.17 & $0.89 [+0.19, -0.17]$ \\
8.00 -- 11.31 & 3.17 -- 3.99 & $0.17 [+0.062, -0.055]$ &
8.00 -- 11.31 & 3.99 -- 5.02 & $0.081 [+0.026, -0.023]$ \\
8.00 -- 11.31 & 5.02 -- 6.32 & $0.018 [+0.011, -0.0088]$ &
8.00 -- 11.31 & 6.32 -- 7.96 & $0.0059 [+0.0053, -0.0036]$ \\
8.00 -- 11.31 & 7.96 -- 10.02 & $0.029 [+0.0076, -0.0066]$ &
8.00 -- 11.31 & 10.02 -- 12.62 & $0.054 [+0.0091, -0.0085]$ \\
8.00 -- 11.31 & 12.62 -- 15.89 & $0.015 [+0.010, -0.0076]$ &
8.00 -- 11.31 & 15.89 -- 20.00 & $0.024 [+0.038, -0.018]$ \\
11.31 -- 16.00 & 2.00 -- 2.52 & $1.2 [+0.52, -0.44]$ &
11.31 -- 16.00 & 2.52 -- 3.17 & $0.97 [+0.27, -0.25]$ \\
11.31 -- 16.00 & 3.17 -- 3.99 & $0.27 [+0.093, -0.082]$ &
11.31 -- 16.00 & 3.99 -- 5.02 & $0.045 [+0.036, -0.027]$ \\
11.31 -- 16.00 & 5.02 -- 6.32 & $0.11 [+0.026, -0.024]$ &
11.31 -- 16.00 & 6.32 -- 7.96 & $0.0095 [+0.010, -0.0064]$ \\
11.31 -- 16.00 & 7.96 -- 10.02 & $0.029 [+0.011, -0.0093]$ &
11.31 -- 16.00 & 10.02 -- 12.62 & $0.063 [+0.014, -0.012]$ \\
11.31 -- 16.00 & 12.62 -- 15.89 & $0.021 [+0.017, -0.012]$ &
11.31 -- 16.00 & 15.89 -- 20.00 & $0.050 [+0.082, -0.038]$ \\
\hline
\end{tabular}
\end{table*}

\begin{table*}
\caption{Occurrence rates (number of planets per 100 stars) of 100\,pc sample of Fig.~\ref{fig:100pc-occurrence}. Values are reported to two significant figures.}
\label{tab:TESS-SPOC 100 pc table}
\renewcommand{\arraystretch}{1.5}
\begin{tabular}{ccc @{\ \vline\ } ccc}
\hline
Period ranges (days) & Radius ranges ($R_{\earth}$) & Occurrence rates (\%)
   & Period ranges (days) & Radius ranges ($R_{\earth}$) & Occurrence rates (\%) \\ 
\hline
0.50 -- 1.00 & 2.00 -- 3.17 & $0.046 [+0.037, -0.025]$ &
0.50 -- 1.00 & 3.17 -- 5.02 & $0.022 [+0.024, -0.014]$ \\
0.50 -- 1.00 & 5.02 -- 7.96 & $0.0098 [+0.015, -0.0073]$ &
0.50 -- 1.00 & 7.96 -- 12.62 & $0.021 [+0.025, -0.014]$ \\
0.50 -- 1.00 & 12.62 -- 20.00 & $0.061 [+0.075, -0.043]$ &
1.00 -- 2.00 & 2.00 -- 3.17 & $0.10 [+0.068, -0.048]$ \\
1.00 -- 2.00 & 3.17 -- 5.02 & $0.017 [+0.026, -0.013]$ &
1.00 -- 2.00 & 5.02 -- 7.96 & $0.010 [+0.018, -0.0078]$ \\
1.00 -- 2.00 & 7.96 -- 12.62 & $0.011 [+0.019, -0.0086]$ &
1.00 -- 2.00 & 12.62 -- 20.00 & $0.034 [+0.055, -0.025]$ \\
2.00 -- 4.00 & 2.00 -- 3.17 & $0.71 [+0.24, -0.19]$ &
2.00 -- 4.00 & 3.17 -- 5.02 & $0.095 [+0.088, -0.055]$ \\
2.00 -- 4.00 & 5.02 -- 7.96 & $0.073 [+0.057, -0.039]$ &
2.00 -- 4.00 & 7.96 -- 12.62 & $0.058 [+0.046, -0.033]$ \\
2.00 -- 4.00 & 12.62 -- 20.00 & $0.080 [+0.085, -0.054]$ &
4.00 -- 8.00 & 2.00 -- 3.17 & $3.8 [+0.65, -0.59]$ \\
4.00 -- 8.00 & 3.17 -- 5.02 & $0.32 [+0.16, -0.13]$ &
4.00 -- 8.00 & 5.02 -- 7.96 & $0.031 [+0.051, -0.023]$ \\
4.00 -- 8.00 & 7.96 -- 12.62 & $0.052 [+0.061, -0.035]$ &
4.00 -- 8.00 & 12.62 -- 20.00 & $0.29 [+0.25, -0.16]$ \\
8.00 -- 16.00 & 2.00 -- 3.17 & $7.0 [+1.2, -1.1]$ &
8.00 -- 16.00 & 3.17 -- 5.02 & $0.32 [+0.28, -0.18]$ \\
8.00 -- 16.00 & 5.02 -- 7.96 & $0.049 [+0.079, -0.037]$ &
8.00 -- 16.00 & 7.96 -- 12.62 & $0.18 [+0.15, -0.10]$ \\
8.00 -- 16.00 & 12.62 -- 20.00 & $1.5 [+0.91, -0.69]$ & \\
\hline
\end{tabular}
\end{table*}

\section{Detailed posterior distribution comparisons}\label{appendix}
Here, selected from Fig.~\ref{fig:post-comp}, we present the four distributions with the greatest discrepancies when different thresholds are applied. 

\begin{figure}
    \centering
    \includegraphics[width=1\columnwidth]{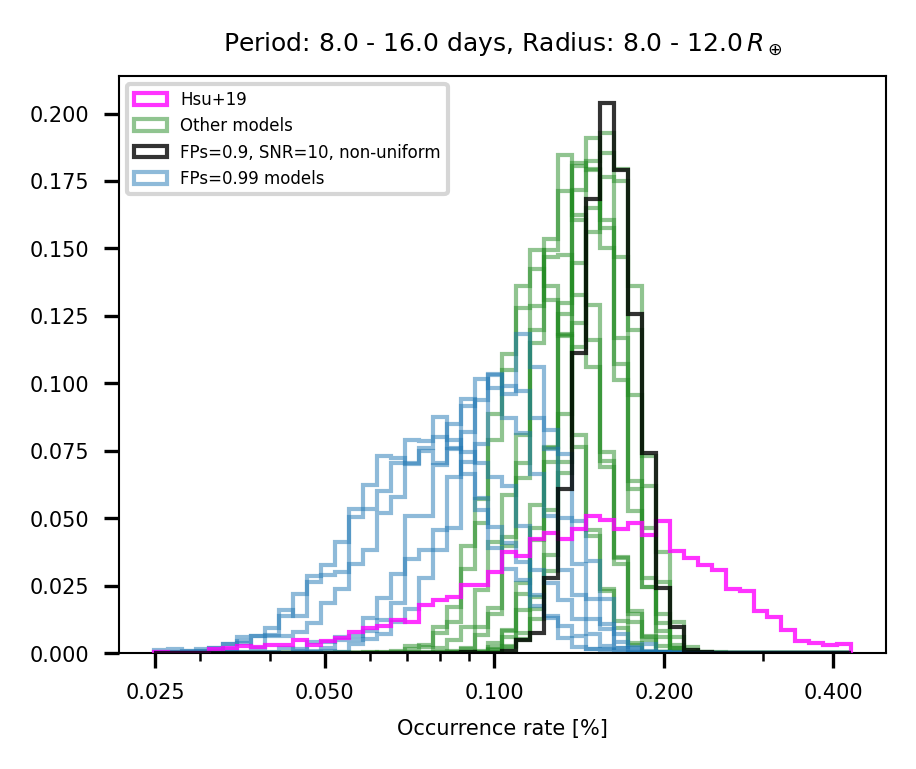}
    \caption{Posterior distributions of occurrence rates for various models and the Kepler results in a cell with an unusually large discrepancy; the cell’s location is indicated in the title. Different colours denote models with different thresholds, as labelled in the legend.}
    \label{fig:8-16d-8-12e}
\end{figure}

\begin{figure}
    \centering
    \includegraphics[width=1\linewidth]{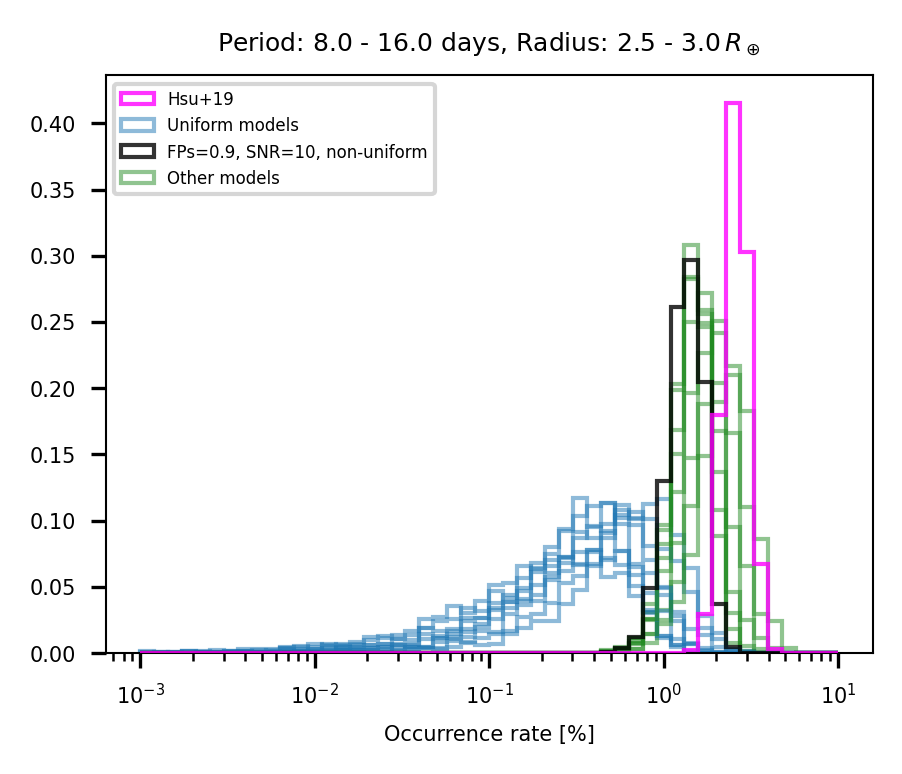}
    \caption{Similar to Fig.~\ref{fig:8-16d-8-12e}.}
    \label{fig:8-16d-2.5-3e}
\end{figure}

\begin{figure}
    \centering
    \includegraphics[width=1\linewidth]{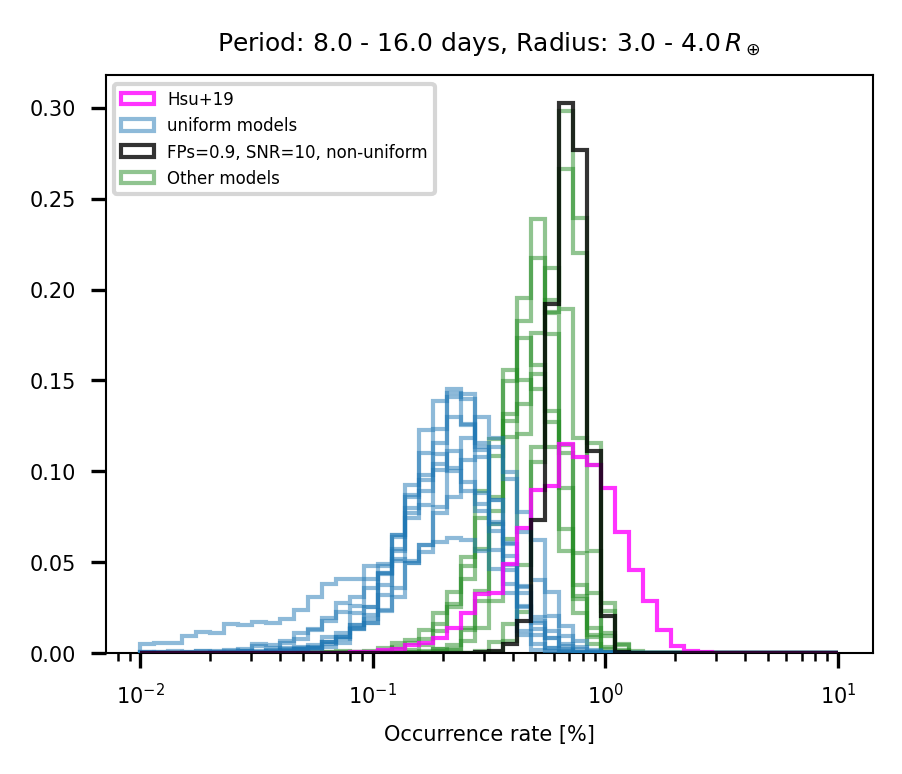}
    \caption{Similar to Fig.~\ref{fig:8-16d-8-12e}.}
    \label{fig:8-16d-3-4e}
\end{figure}

\begin{figure}
    \centering
    \includegraphics[width=1\linewidth]{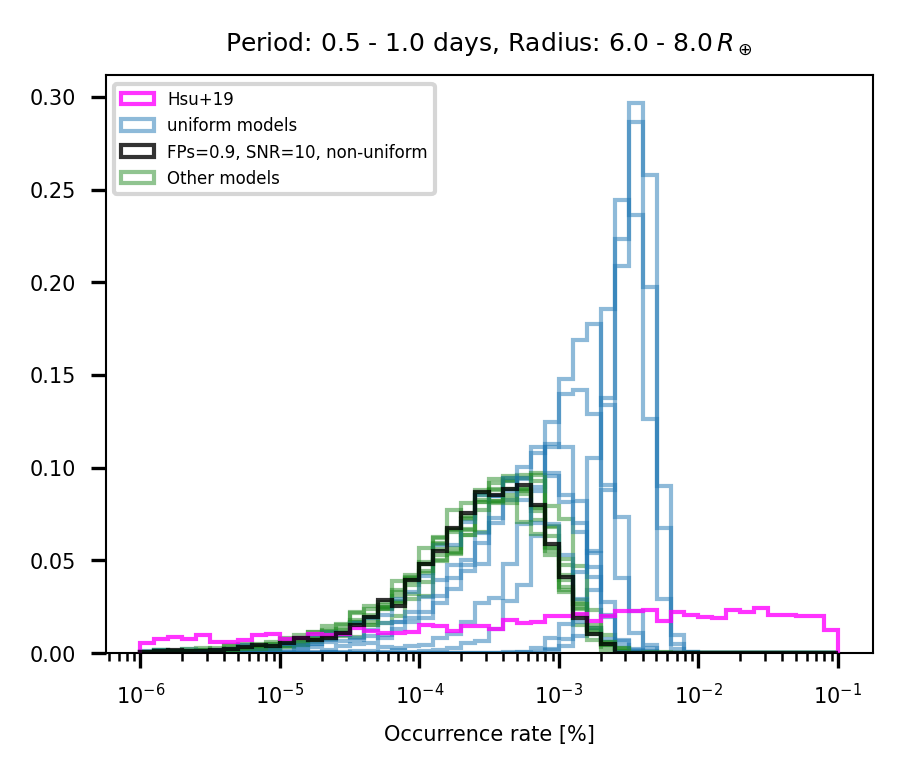}
    \caption{Similar to Fig.~\ref{fig:8-16d-8-12e}.}
    \label{fig:0.5-1d-6-8e}
\end{figure}

Fig.~\ref{fig:8-16d-8-12e} shows that all models with a 0.99 FP threshold have a significantly lower occurrence rate, while others remain generally consistent, indicating that an excessively high FP threshold eliminates too many candidates, while the completeness estimation fail to account for this. Fig.~\ref{fig:8-16d-2.5-3e}, Fig.~\ref{fig:8-16d-3-4e}, and Fig.~\ref{fig:0.5-1d-6-8e} demonstrate that all models trained on uniformly injected planetary samples show a lower occurrence rate, suggesting these samples are highly deficient in that region and leading to the very limited completeness reliability. 

\bsp	
\label{lastpage}
\end{document}